\documentclass[12 pt]{article}
\pdfoutput=1
\usepackage{amsmath,mathrsfs}
\usepackage{amssymb}
\allowdisplaybreaks
\usepackage{amscd}
\usepackage{graphicx}
\usepackage{tabularx,booktabs}
\usepackage{array}
\newcolumntype{Z}{>{\centering\let\newline\\\arraybackslash\hspace{0pt}}X}
\usepackage[footnotesize,bf,textfont={it},margin=1 cm]{caption}

\def\be{\begin{equation}}
\def\ee{\end{equation}}
\def\frc#1#2{\relax\ifmmode{\textstyle\frac{#1}{#2}} 
                    \else$\frac{#1}{#2}$\fi}         

\def\rI{{{}_{\rm I}}}
\def\rJ{{{}_{\rm J}}}

\def\hj{{\hat\jmath}}
\def\hk{{\hat k}}
\def\hi{{\hat\imath}}

\def\fracm#1#2{\hbox{\large{${\frac{{#1}}{{#2}}}$}}}

\def\vCent#1{\vcenter{\hbox{\hss#1\hss}}}

\def\pp{{\mathchoice
          {
              \kern 1pt%
              \raise 1pt
              \vbox{\hrule width5pt height0.4pt depth0pt
                    \kern -2pt
                    \hbox{\kern 2.3pt
                          \vrule width0.4pt height6pt depth0pt
                          }
                    \kern -2pt
                    \hrule width5pt height0.4pt depth0pt}%
                    \kern 1pt
           }
            {
              \kern 1pt%
              \raise 1pt
              \vbox{\hrule width4.3pt height0.4pt depth0pt
                    \kern -1.8pt
                    \hbox{\kern 1.95pt
                          \vrule width0.4pt height5.4pt depth0pt
                          }
                    \kern -1.8pt
                    \hrule width4.3pt height0.4pt depth0pt}%
                    \kern 1pt
            }
            {
              \kern 0.5pt%
              \raise 1pt
              \vbox{\hrule width4.0pt height0.3pt depth0pt
                    \kern -1.9pt  
                    \hbox{\kern 1.85pt
                          \vrule width0.3pt height5.7pt depth0pt
                          }
                    \kern -1.9pt
                    \hrule width4.0pt height0.3pt depth0pt}%
                    \kern 0.5pt
            }
            {
              \kern 0.5pt%
              \raise 1pt
              \vbox{\hrule width3.6pt height0.3pt depth0pt
                    \kern -1.5pt
                    \hbox{\kern 1.65pt
                          \vrule width0.3pt height4.5pt depth0pt
                          }
                    \kern -1.5pt
                    \hrule width3.6pt height0.3pt depth0pt}%
                    \kern 0.5pt
            }
        }}

\def\mm{{\mathchoice
                       {
                             \kern 1pt
               \raise 1pt    \vbox{\hrule width5pt height0.4pt depth0pt
                                  \kern 2pt
                                  \hrule width5pt height0.4pt depth0pt}
                             \kern 1pt}
                       {
                            \kern 1pt
               \raise 1pt \vbox{\hrule width4.3pt height0.4pt depth0pt
                                  \kern 1.8pt
                                  \hrule width4.3pt height0.4pt depth0pt}
                             \kern 1pt}
                       {
                            \kern 0.5pt
               \raise 1pt
                            \vbox{\hrule width4.0pt height0.3pt depth0pt
                                  \kern 1.9pt
                                  \hrule width4.0pt height0.3pt depth0pt}
                            \kern 1pt}
                       {
                           \kern 0.5pt
             \raise 1pt  \vbox{\hrule width3.6pt height0.3pt depth0pt
                                  \kern 1.5pt
                                  \hrule width3.6pt height0.3pt depth0pt}
                           \kern 0.5pt}
                       }}

\def\ad{{\kern0.5pt
                   \alpha \kern-5.05pt \raise5.8pt\hbox{$\textstyle.$}\kern
0.5pt}}

\def\bd{{\kern0.5pt
                   \beta \kern-5.05pt \raise5.8pt\hbox{$\textstyle.$}\kern
0.5pt}}

\def\qd{{\kern0.5pt
                   q \kern-5.05pt \raise5.8pt\hbox{$\textstyle.$}\kern
0.5pt}}
\def\Dot#1{{\kern0.5pt
     {#1} \kern-5.05pt \raise5.8pt\hbox{$\textstyle.$}\kern
0.5pt}}

\catcode`@=11
\def\un#1{\relax\ifmmode\@@underline#1\else
        $\@@underline{\hbox{#1}}$\relax\fi}
\catcode`@=12

\def\a{\alpha}
\def\b{\beta}

\def\d{\delta}

\def\l{\lambda}

\def\t{\tau}

\def\dslash{\not{\hbox{\kern-2pt $\partial$}}}
\def\Dslash{\not{\hbox{\kern-4pt $D$}}}
\def\pslash{\not{\hbox{\kern-2.3pt $p$}}}
 \newtoks\slashfraction
 \slashfraction={.13}
 \def\slash#1{\setbox0\hbox{$ #1 $}
 \setbox0\hbox to \the\slashfraction\wd0{\hss \box0}/\box0 }
 
\font\ro=cmsy10                          
\def\kcr{{\hbox{\ro \char'170}}}                
\def\ktl{{\hbox{\ro \char'170}}}        
\def\ktr{{\hbox{\ro \char'170}}}        
\def\kbl{{\hbox{\ro \char'170}}}        
\def\kbr{{\hbox{\ro \char'170}}}        

\def\plpl{\raise-2pt\hbox{$\raise3pt\hbox{$_+$}\hskip-6.67pt\raise0.0pt
\hbox{$^+$}\hskip 0.01pt$}}
\def\mimi{\raise-2pt\hbox{$\raise3pt\hbox{$_-$}\hskip-6.67pt\raise0.0pt
\hbox{$^-$}\hskip 0.01pt$}} 

\def\bo{{\raise.15ex\hbox{\large$\Box$}}}               
\def\pa{\partial}                                       
\def\TH{{\raise.2ex\hbox{$\displaystyle \bigodot$}\mskip-4.7mu \llap H \;}}
\def\face{{\raise.2ex\hbox{$\displaystyle \bigodot$}\mskip-2.2mu \llap {$\ddot
        \smile$}}}                                      

\def\Tilde#1{\widetilde{#1}}                    
\def\Hat#1{\widehat{#1}}                        
\def\leftrightarrowfill{$\mathsurround=0pt \mathord\leftarrow \mkern-6mu
        \cleaders\hbox{$\mkern-2mu \mathord- \mkern-2mu$}\hfill
        \mkern-6mu \mathord\rightarrow$}
\def\dvec#1{\vbox{\ialign{##\crcr
        \leftrightarrowfill\crcr\noalign{\kern-1pt\nointerlineskip}
        $\hfil\displaystyle{#1}\hfil$\crcr}}}           
\def\dt#1{{\buildrel {\hbox{\LARGE .}} \over {#1}}}     

\def\fracm#1#2{\hbox{\large{${\frac{{#1}}{{#2}}}$}}}
\def\frac#1#2{{\textstyle{#1\over\vphantom2\smash{\raise.20ex
        \hbox{$\scriptstyle{#2}$}}}}}                   
\def\sfrac#1#2{{\vphantom1\smash{\lower.5ex\hbox{\small$#1$}}\over
        \vphantom1\smash{\raise.4ex\hbox{\small$#2$}}}} 
\def\bfrac#1#2{{\vphantom1\smash{\lower.5ex\hbox{$#1$}}\over
        \vphantom1\smash{\raise.3ex\hbox{$#2$}}}}       
\def\afrac#1#2{{\vphantom1\smash{\lower.5ex\hbox{$#1$}}\over#2}}    
\def\on#1#2{\mathop{\null#2}\limits^{#1}}               

\def\pa{\partial}

\def\ad{{\dot{\alpha}}}
\def\bd{{\dot{\beta}}}

\font\ro=cmsy10                          
\def\kcr{{\hbox{\ro \char'170}}}                
\def\ktl{{\hbox{\ro \char'170}}}        
\def\ktr{{\hbox{\ro \char'170}}}        
\def\kbl{{\hbox{\ro \char'170}}}        
\def\kbr{{\hbox{\ro \char'170}}}        

\parskip=\medskipamount                 
\thicklines                         

\def\border{                                            
        \setlength{\unitlength}{1mm}
        \newcount\xco
        \newcount\yco
        \xco=-21
        \yco=12
        \begin{picture}(140,0)
        \put(\xco,\yco){$\ktl$}
        \advance\yco by-1
        {\loop
        \put(\xco,\yco){$\kcr$}
        \advance\yco by-2
        \ifnum\yco>-240
        \repeat
        \put(\xco,\yco){$\kbl$}}
        \xco=158
        \yco=12
        \put(\xco,\yco){$\ktr$}
        \advance\yco by-1
        {\loop
        \put(\xco,\yco){$\kcr$}
        \advance\yco by-2
        \ifnum\yco>-240
        \repeat
        \put(\xco,\yco){$\kbr$}}
        \put(-20,13){\tiny **University of Maryland * Center for String and
         Particle  Theory* Physics Department***University of Maryland *Center
        for String and Particle  Theory** }
        \put(-20,-241.5){\tiny **University of Maryland * Center for String and
         Particle  Theory* Physics Department***University of Maryland *Center
        for String and Particle  Theory** }
        \end{picture}
        \par\vskip-8mm}

\def\headpic{                                           
        \indent
        \setlength{\unitlength}{.4mm}
        \thinlines
        \par
        \begin{picture}(29,16)
        \put(165,16){\line(1,0){4}}
        \put(170,16){\line(1,0){4}}
        \put(180,16){\line(1,0){4}}
        \put(175,0){\line(1,0){4}}
        \put(180,0){\line(1,0){4}}
        \put(185,0){\line(1,0){4}}
        \put(169,0){\line(0,1){16}}
        \put(170,0){\line(0,1){16}}
        \put(179,0){\line(0,1){16}}
        \put(180,0){\line(0,1){16}}
        \put(184,0){\line(0,1){16}}
        \put(185,0){\line(0,1){16}}
        \put(169,16){\oval(8,32)[bl]}
        \put(170,16){\oval(8,32)[br]}
        \put(179,0){\oval(8,32)[tl]}
        \put(185,0){\oval(8,32)[tr]}
        \end{picture}
        \par\vskip-6.5mm
        \thicklines}
\def\endtitle{\end{quotation}\newpage}                  

\newskip\humongous \humongous=0pt plus 1000pt minus 1000pt
\def\caja{\mathsurround=0pt}
\def\eqalign#1{\,\vcenter{\openup2\jot \caja
        \ialign{\strut \hfil$\displaystyle{##}$&$
        \displaystyle{{}##}$\hfil\crcr#1\crcr}}\,}
\newif\ifdtup

\begin{document}

\def\dt#1{\on{\hbox{\bf .}}{#1}}                
\def\Dot#1{\dt{#1}}

\def\gfrac#1#2{\frac {\scriptstyle{#1}}
        {\mbox{\raisebox{-.6ex}{$\scriptstyle{#2}$}}}}
\def\gg{{\hbox{\sc g}}}
\border\headpic {\hbox to\hsize{
\hfill
{UMDEPP-015-003}}}
\par \noindent
{ \hfill
}
\par

\setlength{\oddsidemargin}{0.3in}
\setlength{\evensidemargin}{-0.3in}
\begin{center}
{\large\bf Think Different: Applying the Old Macintosh Mantra to the
\\ [.075in]Computability of the SUSY Auxiliary Field Problem}\\[.3in]
Mathew Calkins\footnote{mathewpcalkins@gmail.com}${}^{\dagger}$, D.\ E.\ A.\ 
Gates\footnote{deagates@terpmail.umd.edu}${}^{\dagger}$, S.\, 
James Gates, Jr.\footnote{gatess@wam.umd.edu}${}^{\dagger}$, and \\
William M. Golding\footnote{william.m.golding2.civ@mail.mil}${}^{*}$
\\[0.4in]
${}^\dag${\it Center for String and Particle Theory\\
Department of Physics, University of Maryland\\
College Park, MD 20742-4111 USA} 
\\[0.1in]
and
\\[0.1in]
${}^*${\it Sensors and Electron Devices Directorate\\
US Army Research Laboratory\\
Adelphi, Maryland 20783 USA} 
\\[0.6in]
{\bf ABSTRACT}\\[.01in]
\end{center}
\begin{quotation}
{Starting with valise supermultiplets obtained from 0-branes plus 
field redefinitions, valise adinkra networks, and the ``Garden Algebra,'' 
we discuss an architecture for algorithms that (starting from on-shell 
theories and, through a well-defined computation procedure), search
for off-shell completions.  We show in one dimension how to directly 
attack the notorious ``off-shell auxiliary field'' problem of supersymmetry 
with algorithms in the adinkra network-world
formulation.}
\\[0.2in]
\vfill
\endtitle

\setcounter{equation}{0}
\section{Introduction}
$~~~$ 
In 1979, one of the authors (SJG) was invited to the California
Institute of Technology by Dr.\ J.\ H.\ Schwarz for a program of study on the issue
of finding a set of auxiliary fields with which to close the supersymmetry algebra 
on the component fields of the 10D, $\cal N$ = 1 Maxwell vector supermultiplet without
the use of equations of motion.  The study was not completely satisfactory
as no set of such fields were identified.  This situation has remains 
unchanged.

Later in 1981, there was formulated a ``No-Go'' theorem \cite{SRThm} which apparently 
explained the result of the earlier study.  The abstract to this paper by Siegel and 
Ro\v cek that presented the theorem read

$~~~~~~$ {\it{Applying a simple counting argument to all supermultiplets, we find}}
\newline \noindent $~~~~~~$ 
{\it {that for the
$\cal N$ = 4 super Yang-Mills theory the auxiliary field problem}}
\newline \noindent $~~~~~~$ 
{\it {cannot have a solution 
within any previously known framework.  We pro-}}
\newline \noindent $~~~~~~$ 
{\it {pose alternatives.}} \vskip0.05in
\noindent
Since the 4D, $\cal N$ = 4 Maxwell vector supermultiplet  is related to the
10D, $\cal N$ = 1 version via torus compactification, the study
result would seem covered.  

While this argument is simple and elegant, it has at least one puzzling aspect.

It is widely accepted that the 10D, $\cal N$ = 1 and 4D, $\cal N$ = 4 Maxwell 
supermultiplets can be embedded within a formulation involving unconstrained 
super $p$-forms.   This fact should imply the existence of some type of off-shell 
formulation containing the fields of the on-shell theory.  Based on this superspace 
argument there should exist an off-shell completion of the 4D, $\cal N$ = 4 Maxwell 
supermultiplet.  

It has been known since the work of \cite{GV} that the super 1-form formulation 
of the 10D, $\cal N$ = 1 Maxwell and Yang-Mills supermultiplets have a rather 
distinctive structure in terms of the constraints that describe the theories in 
comparison to other similar theories.  In the 10D case, there is a spinor-spinor 
field strength component that vanishes in three, four, and six dimensions
\cite{BRSz} is non-zero in ten dimensions.  This difference was used in 
the work of  \cite{GV} to provide the first superspace description of the lowest 
order open superstring corrections and has been verified a number of times 
since (see e.\ g.\ \cite{BRSz,PVW,CNT,K,BP}).

Thus, there exists a contrast between these two widely accepted results.  The 
work in \cite{SRThm} concerns the action (with off-shell supersymmetry), while 
\cite{GV} gives only the field equations (thus on-shell, although with contributions 
from integrating out higher massive modes from the open superstring. 
Of course, the resolution of the contrast must lie in the fact that some assumption 
made in one approach does not apply to the other.  Knowing this, however, does 
not provide a detailed explanation.

In 1995, the presence of sets of matrices with certain regularities \cite{ENUF3,ENUF4} 
was noted to occur (presumably) in all 1D systems that realize supersymmetry in 
a linear manner.  The matrices (given the designation of L-matrices and R-matrices)
would later become recognized as the adjacency matrices of adinkra networks
\cite{adnk1}.  This latter identification became critical in providing a definition of 
these matrices independent of field theory models and opening a path to totally 
unexpected connections to subjects such as cubical cohomology \cite{DFGHILM4}, 
error-correcting codes \cite{DFGHILM,DFGHILM2,DFGHILM3}, ranked poset
\cite{posets}, Coxeter Groups 
\cite{permutadnk2} and most recently Riemann surfaces \cite{adnkgeo}.

Soon after our discovery of the ubiquity of L-matrices and R-matrices in 1D SUSY
 theories, we proposed \cite{ENUF,ENUF2,ENUF5} that these might play a vital role 
in attacking the off-shell SUSY auxiliary field problem via a technique given the acronym
of RADIO.  It was envisioned that this technique permits the derivation of new 
on-shell and off-shell representations by starting from a D-dimensional, $\cal N$-extended 
theory. The steps of the process begin by reducing (R) the higher dimensional theory 
to 1D, followed by performing certain Òautomorphic dualityÓ (AD) transformations, next
integrating additional 1D representations (I) and then oxidizing (O) back up to
the higher dimensional spacetime.  The actions of (R) and (AD) together produce 
what we now refer to as ``valise supermultiplets.''  These provide the starting
point in this current work.  The main focus of this work
will be to indicate how the (I) can be carried out.  We discuss the general
philosophy of this step and also show by explicit calculation how this is done.

\section{Review of the Siegel-Ro\v cek Theorem} 
$~~~$
During 2009 in exchanges between M. Faux and SJG,  
the following discussion for understanding the essential points of the 
Siegel-Ro\v cek argument were noted.

The smallest off-shell $\cal N$-extended supermultiplets in four-dimensions 
have $2^{2N-1}$ component fermions and the same number of component 
bosons.  The number of off-shell component fermions $F$ in any non-minimal 
4D supermultiplet must be an integer multiple of the component fermions in 
the minimal multiplet. Thus, $F$ = $2^{2N-1}\, m$, where m is some positive 
integer.

All fermions carry an odd number of SO(N) fundamental indices.  As a 
result, all fermions carry an integer multiple of 4$\cal N$ off-shell degrees 
of freedom, where the 4 reflects the dimensionality of a minimal spinor.

A given off-shell multiplet has $f$ fermionic degrees of freedom 
corresponding to propagating degrees of freedom, plus some number of 
auxiliary fermions. Auxiliary fermions come paired. It follows that $F$ = 
$f$ + $2 ( 4{\cal N})\, n$, where $n$ counts the number of auxiliary fermion
reduced pairings.  Thus, $n$ counts the number of minimal spinors assembled 
to form a given auxiliary fermion representation.

By comparing the two restrictions on the number of fermion components, 
we conclude $f$ + 8$\cal N$ $n$ = $2^{2{\cal N} -1} \,m$.  Adapting this 
to the 4D cases $\cal N$ = 2 and $\cal N$ = 4, this yields

$~~~~$ $\cal N$ = 2 : $f$ + 16 $n$ = 8 $m$ ~~~~~~,

$~~~~$ $\cal N$ = 4 : $f$ + 32 $n$ = 128 $m$ ~~~.
\vskip.012in \noindent
For the cases of the $\cal N$ = 2 vector and tensor supermultiplets, we have 
$f$ = 8, $n$ = 0, and $m$ = 1.  For the case of the $\cal N$ = 4 vector 
supermultiplet, we have $f$ = 16, since there are four physical fermions 
transforming as a 4 under SO(4). We can then rearrange the second equation 
above to read $n$ = 4$m$ $-$ $\fracm 12$.  This equation has no solutions 
for integer $m$ and integer $n$.

\section{In the World of 0-Brane Valise Supermultiplets} 
$~~~$
All our previous explorations suggests that via the (R) and (AD) steps of the
RADIO proposal, any linear representation of spacetime supersymmetry can be
made to depend on a single real parameter in the forward 
light-cone \cite{redux}.  Under field redefinition using derivatives or 
integrals, such representations can be brought to a universal form of a valise 
supermultiplet
\be \eqalign{
{\rm D}{}_a \, \Phi{}_{\Lambda}  ~=~ i\,  \left( {\rm L}{}_{\Lambda}\right){}_{a}
{}^{\Hat {\Lambda}} \, \, \Psi{}_{\Hat {\Lambda}}
~~~,~~~
{\rm D}{}_{a} \Psi{}_{\Hat {\Lambda}} ~=~  \left( {\rm R}{}^{\Lambda} \right)
{}_{\Hat {\Lambda}}{\,}_{a} \, \pa_{\tau} \, \Phi{}_{\Lambda}  ~~~.
} \label{covVAL}
\ee 
Here the explicit forms of the constants $ \left( {\rm L}{}_{\Lambda}\right){
}_{a}{}^{\Hat {\Lambda}}$ and $\left( {\rm R}{}^{\Lambda} \right){}_{\Hat {\Lambda
}}{\,}_{a}$, as well as the field variables $\Phi{}_{\Lambda} $, and $\Psi{}_{\Hat {
\Lambda}}$ vary from supermultiplet to supermultiplet.   The condition that the 
field variables in (\ref{covVAL}) form representations of spacetime supersymmetry 
just takes the simplified form
\be
\{ \,  {\rm D}{}_{a} ~,~  {\rm D}{}_{b} \, \} ~=~ i \, 2 \, (\gamma^0){}_{a \, b}  \pa_{\t}  ~~~~,
\label{SUSYeq}
\ee
when calculated on any of the component fields from any of the supermultiplets. 

Implementing the (R) and (AD) parts of the RADIO proposal
necessarily breaks SO(1,3) covariance.  However, as noted in \cite{G-1}, in place of
the SO(1,3) symmetry a new SU(2) $\otimes$ SU(2) symmetry appears in the
equations that emerge for off-shell valise supermultiplets.  The 
generators of these two commuting SU(2) symmetries are given by
\be
i \, \fracm 14
[\gamma^m,\, \gamma^n]    ~~~,
\label{su2a}
\ee
for the generator of purely spatial rotations and
\be
 i \gamma^0 ~~,~~  \gamma^5 ~~,~~  \gamma^0 \gamma^5  ~~~,
\label{su2b}
\ee
for the generators of an extended SU(2) R-symmetry  \cite{adnkholor1}.

To make this more concrete, we illustrate some familiar representations after 
application of the (R) and (AD) steps and obtain the results of \cite{G-1}: \newline
\noindent
(a.) Chiral Supermultiplet (CS);
\be
\eqalign{
{\rm D}_a A  ~&=~ \psi_a  ~~,~~
{\rm D}_a B   ~=~ i \, ( \gamma^5 )_a{}^b \psi_b  ~~,~~  
{\rm D}_a F  ~=~ ( \gamma^0)_a{}^b \, \psi_b  ~~, ~~
{\rm D}_a G   ~=~ i\, ( \gamma^5 \gamma^0 )_a{}^b \, \psi_b  ~~, 
\cr
{\rm D}_a \psi_b  ~&=~ i\, ( \gamma^0 )_{ab}  \left( \,\partial_{
\tau} A  \, \right) - ( \gamma^5 \gamma^0)_{ab}  \left( \, 
\partial_{\tau} B  \, \right) ~-~ i C_{ab}  \left( \,  \partial_{\tau} F  \, \right) + 
( \gamma^5 )_{ab}  \left( \, \partial_{\tau} G   \, \right) 
~~~,
} \label{BV2ZZ} 
\ee
\noindent
(b.) Vector Supermultiplet (VS);
\be
\eqalign{
{\rm D}{}_a A_m & ~=~ (\gamma_m)_a{}^b \lambda_b  ~~~, ~~~ {\rm D}_a 
{\rm d}  \, =\,  i (\gamma^5 \gamma^0)_a{}^b \, \lambda_b 
 ~~~~~~\,~~~~, \cr
{\rm D}{}_a \lambda_b & ~=~ - i \, (  \gamma^0 
\gamma^m )_{ab} \,  \left( \,  \pa_{\tau} A_m  \, \right)~+~ (\gamma^5
)_{ab} \,  \left( \, \pa_{\tau} {\rm d} \, \right) 
~~~,  
}  \label{BV4ZZ} 
\ee
\noindent
(c.) Tensor Supermultiplet (TS);
\be
\eqalign{
{\rm D}{}_a \varphi ~&=~ \chi_a   ~~~, ~~~
{\rm D}{}_a B_{m \, n}  ~=~ - \tfrac{1}{4} ([\gamma_m,\, 
\gamma_n])_a{}^b \chi_b  ~\,~~, \cr
{\rm D}{}_a \chi_b ~&=~ i (\gamma^0)_{ab} \,  \partial_{\tau}
\varphi - i  \tfrac{1}{2} (\gamma^0 \, [\gamma^m,\, \gamma^n]
)_{ab}  \, \partial_\tau B_{m \, n}  
 ~~~\,~,
}  \label{BV6ZZ} 
\ee
\noindent
(d.) Axial vector Supermultiplet (AVS); 
\be  \eqalign{
{\rm D}{}_a U_{m} & ~=~ i \, (\gamma^5 \gamma_{m})_a{}^b {\Tilde 
{\lambda}}{}_b   ~~,~~
 {\rm D}_a {\Tilde {\rm d}}  ~=~  -\,  ( \gamma^{0})_a{}^b \, \pa_{\tau} 
{\Tilde {\lambda}}{}_b  ~~,  
 \cr
{\rm D}{}_a {\Tilde {\lambda}}{}_b & ~=~   (\gamma^5  \gamma^{0}
\gamma^{m} )_{ab} \,   \left( \,  \pa_{\tau} U_m  \, \right)  \, ~+~ i C{}_{ab} \,   
{\Tilde {\rm d}} ~~,~~}  \label{BVa4ZZ}
\ee
\noindent
(e.) Axial tensor Supermultiplet (ATS); and
\be
\eqalign{
{\rm D}{}_a {\Tilde \varphi} ~&=~   i \, ( \gamma^5 )_a{}^b  {\Tilde \chi}{}_b   
~~~, ~~~ {\rm D}{}_a  {\Tilde B}{}_{m \, n}  ~=~ -i\,  \tfrac{1}{4} (  \gamma^5 
[\gamma_m,\, \gamma_n])_a{}^b  {\Tilde \chi}{}_b  ~\,~~, \cr
{\rm D}{}_a  {\Tilde \chi}{}_b ~&=~ - (\gamma^0 \gamma^5)_{ab} \,  \partial_{
\tau}{\Tilde \varphi} +   \tfrac{1}{2} (\gamma^0 \gamma^5 \, [\gamma^m,\, 
\gamma^n])_{ab}  \, \partial_\tau  {\Tilde B}{}_{m \, n}   
 ~~~~~~~~~~~~.
}  \label{BV6ZZz} 
\ee
\noindent
(f.) Real Scalar Supermultiplet (RSS);
\be
\eqalign{
{\rm D}{}_{a} K & = \zeta _a ~,~ 
{\rm D}{}_{a} {\rm d}  = - \left( \gamma ^0 \right) _a ^{\ d} \Lambda _d 
~~~, \cr
{\rm D}{}_{a} M  &
= \frac{1}{2} \Lambda _a - \frac{1}{2} \left( \gamma^0 \right)_a^{
\ d} \zeta _d ~,~  
{\rm D}{}_{a} N  = - i \frac{1}{2} \left( \gamma^5 \right) _a ^{\ d} \Lambda _d + i 
\frac{1}{2} \left( \gamma ^5 \gamma ^0 \right) _a ^{\ d} \zeta _d  ~~, \cr
{\rm D}{}_{a} U_0 & = i \frac{1}{2} \left( \gamma^5 \gamma _0 \right) _a ^{\ d} 
\Lambda_d - i \frac{1}{2} \left( \gamma^5 \right) _a ^{ \ d} \zeta _d 
~,~
{\rm D}{}_{a} U_m  = i \frac{1}{2} \left( \gamma^5 \gamma _m \right) _a ^{\ d} 
\Lambda _d - i \frac{1}{2} \left( \gamma^5 \gamma^0 \gamma_m 
\right)_a ^{ \ d} \zeta _d ~,~ \cr
{\rm D}{}_{a} \zeta _b & = i \left( \gamma ^0 \right) _{ab}  \pa_{\tau}K + \left( 
\gamma^5 \gamma^\mu \right) _{ab} \pa_{\tau}U_\mu  + i C_{ab} \pa_{
\tau}M  + \left( \gamma^5 \right) _{ab} \pa_{\tau}N ~~~,  \cr
{\rm D}{}_{a} \Lambda _b & = i \left( \gamma ^0 \right) _{ab}  \pa_{\tau}M  + \left( 
\gamma^5 \gamma^0 \right) _{ab} \pa_{\tau}N + \left( \gamma^5 \gamma^0 
\gamma^\nu \right) _{ab} \pa_{\tau}U_\nu  + i C_{ab} \pa_{\tau}{\rm d}  ~~~.
}
\label{R1}
\ee

In particular for each of the supermultiplets, one can define a `vector' of
bosonic (denoted by $\Phi{}_{\Lambda}$) and fermionic (denoted by 
$\Psi{}_{\Hat \Lambda}$) valise supermultiplet variables.  In the case
 of the CS we have
\be \eqalign{ {~~~}
\Phi{}_{\Lambda} ~&=~    \left(\, A, \, B, \, F, \, G  \,\right)   ~~,~~~
\Psi{}_{\Hat {\Lambda}} ~=~ \left( \, \psi{}_a  \,\right)   ~~~~~, \cr
} \label{xhi} \ee
for the VS we have
\be \eqalign{ {~~~~}
\Phi{}_{\Lambda} ~&=~     \left(\, A{}_m, \, {\rm d}  \,\right)   ~~~~~\,~~~,~~~~
\Psi{}_{\Hat {\Lambda}} ~=~ \left( \, \lambda{}_a  \,\right)   ~~~~~, \cr
} \label{vtR} \ee
for the TS we have
\be \eqalign{ {~~~~~}
\Phi{}_{\Lambda} ~&=~     \left(\, \varphi , \, B{}_{m \,n}   \,\right)   ~~~~~\,~,~~~~
\Psi{}_{\Hat {\Lambda}} ~=~ \left( \, \chi{}_a  \,\right)   ~~~~~, \cr
} \ee
for the AVS we have
\be \eqalign{ {~~~~~}
\Phi{}_{\Lambda} ~&=~     \big(\, U{}_m, \, {\Tilde {\rm d}}  \,\big)   ~~~~~~~,~~~~
\Psi{}_{\Hat {\Lambda}} ~=~ \big( \,  {\Tilde {\lambda}}{}_a  \,\big)   ~~~~~\,~, \cr
} \ee
for the ATS we have
\be \eqalign{ {~~~~~}
\Phi{}_{\Lambda} ~&=~     \left(\, {\Tilde \varphi} , \, {\Tilde B}{}_{m \,n}   \,\right)   
~~~~~\,~,~~~~ \Psi{}_{\Hat {\Lambda}} ~=~ \left( \, {\Tilde \chi}{}_a  \,\right)   
~~~~~, \cr
} \ee
and for the RSS we have
\be \eqalign{ {~~~~~}
\Phi{}_{\Lambda} ~&=~     \left(\, {\rm K} , \, M, \, N, \, U_0 , \, U_m , \, {\rm d}  \,\right)   
~~~~~\,~,~~~~ \Psi{}_{\Hat {\Lambda}} ~=~ \left( \, {\zeta}{}_a , \, 
{\Lambda}{}_a  \,\right)   
~~~~~. \cr
} \label{xhiZ} \ee
As seen above, the $\Lambda$ indices are allowed to range over distinct bosonic
representations of SO(1,3) and similarly the $\Hat \Lambda$ indices (in the most 
general case) are allowed to range over distinct fermionic representations of 
SO(1,3).

The explicit forms of the ${\rm L}{}_{\Lambda}$ and ${\rm R}{}^{\Lambda}$ coefficients 
can now be read out for each of the supermultiplets.   Furthermore, as seen 
from these examples, the ${\rm L}{}_{\Lambda}$ and ${\rm R}{}^{\Lambda}$ coefficients 
are constructed from Lorentz invariant tensors, $\gamma$-matrices, and powers thereof.  
Thus, information about the space-time spin of the fields in the supermultiplets is encoded 
in these coefficients even though the field variables only depend on time.  We conjecture 
that every linear off-shell representation of supersymmetry can always be subject to 
0-brane reduction (R), field redefinitions (AD, and possibly linearization) such that equations 
(\ref{covVAL}) and (\ref{SUSYeq}) are satisfied. 

Now in order to focus on the SUSY auxiliary field problem, we concentrate
solely on the chiral supermultiplet and vector supermultiplet in the remainder
of this chapter.  The on-shell version of these two supermultiplets 
are given below.  For the on-shell version of the chiral supermultiplet  we have
\be \eqalign{
{\rm D}_a A ~&=~ \psi_a  ~~~, ~~~~
{\rm D}_a B ~=~ i \, (\gamma^5){}_a{}^b \, \psi_b  ~~~, \cr
{\rm D}_a \psi_b ~&=~ i\, (\gamma^0){}_{a \,b}\,  \partial_{\tau} A 
~-~  (\gamma^5\gamma^0){}_{a \,b} \, \partial_{\tau} B   ~~~,} \label{chi3}
\ee
leading to
\be \eqalign{  {~~~~~}
\{ ~ {\rm D}_a  \,,\,  {\rm D}_b ~\} \, A 
~&=~  i\, 2 \, (\gamma^0){}_{a \,b}\,  \partial_{\tau} \,  A ~~~, ~~~
\{ ~ {\rm D}_a  \,,\,  {\rm D}_b ~\} \, B 
~=~  i\, 2 \, (\gamma^0){}_{a \,b}\,  \partial_{\tau} \, B ~~~, \cr
\{ ~ {\rm D}_a  \,,\,  {\rm D}_b ~\} \, \psi{}_{c}  
~&=~  i\, 2 \, (\gamma^0){}_{a \,b}\,  \partial_{\tau} \,  \psi{}_{c}    
~-~   i \, (\gamma^\mu){}_{a \,b}\, (\gamma_\mu
\gamma^0){}_c{}^d  \partial_{\tau} \,  \psi{}_{d}     ~~~.
} \label{chi4}
\ee
The final term in (\ref{chi4}) is characteristic of an on-shell theory,
an extra term appears relative to the off-shell result shown in
(\ref{SUSYeq}).   Note that (\ref{chi3}) is exactly of the same form
as (\ref{xhi}), but with the important exception that the $F$ and
$G$ field variables are deleted.  In the on-shell theory, the
absence of these two bosonic fields leads to the extra term in
the evaluation of the algebra acting on the fermionic field.  
Going from on-shell to off-shell corresponds by augmenting 
the bosonic vector from $\left( A, \, B  \,\right)$ to $\left( A, \, B, \, F, \, G  \,\right)$
and ensures the condition in (\ref{SUSYeq}) is satisfied.

The on-shell 0-brane formulation of the vector supermultiplet is 
given by
\be \eqalign{
{\rm D}_a \, A{}_{m} ~&=~  (\gamma_m){}_a {}^b \,  \l_b  ~~~, \cr
{\rm D}_a \l_b ~&=~    - i \, (  \gamma^0 
\gamma^m )_{ab} \,  \left( \,  \pa_{\tau} A_m  \, \right)
  ~~~, \cr
} \label{V4}
\ee
and once again we calculate the anticommutator on the fields
\be \eqalign{
{~~~~~~~~~~~~~}
\{ ~ {\rm D}_a  \,,\,  {\rm D}_b ~\} \, A{}_{m}  
~&=~  i\, 2 \, (\gamma^0){}_{a \,b}\,  \partial_\tau \, A{}_{m}  
~~,    \cr
\{ ~ {\rm D}_a  \,,\,  {\rm D}_b ~\} \,\l{}_c
~&=~  i\, 2 \, (\gamma^0){}_{a \,b}\,  \partial_{\tau} \,  \l{}_c   ~-~ i \,
\fracm 12 \,  (\gamma^\mu){}_{a \,b}\,  (\gamma_\mu  
\gamma^0){}_c{}^d \,   \partial_\tau \,  \l{}_d   \cr
&~~~~+~ i \, \fracm 1{16} \,  ([ \, \gamma^\a \, , \, \gamma^\b \,]){}_{a \,b}\,  
( [ \, \gamma_\a \, , \, \gamma_\b \,] \gamma^0){}_c{}^d \,   \partial_\tau \,  \l{}_d   
~~~~~,~
} \label{V5}
\ee
to see the emergence of two extra terms appearing relative to the off-shell'
result shown in (\ref{SUSYeq}).   This review has now set the stage for a statement 
of the off-shell SUSY auxiliary field problem we study in this work.
The result in (\ref{V4}) is the same as the result in (\ref{vtR}) with the exception 
that the latter does not include the d bosonic field variable.  In the on-shell'
theory, the absence of the bosonic d field leads to the extra two terms in the 
evaluation of the algebra acting on the fermionic fields.  So going from on-shell 
to off-shell corresponds to increasing the bosonic vector from $\left( A_m \right)$ 
to $\left( A_m, \, {\rm d}  \, \right)$.

For the 0-brane valise chiral supermultiplet with field content vectors described 
by (\ref{xhi}), the commutator algebra (\ref{SUSYeq}) is satisfied on $\Phi_{
\Lambda}$ and on $\Psi_{\Hat \Lambda}$.  For the 0-brane valise chiral 
supermultiplet with field content vectors described by (\ref{chi3}), the commutator 
algebra (\ref{chi4}) is satisfied on $\Phi_{\Lambda}$ and on $\Psi_{\Hat \Lambda}$.  
For the 0-brane valise vector supermultiplet with field content vectors 
described by (\ref{vtR}), the commutator algebra (\ref{SUSYeq}) is satisfied 
on $\Phi_{\Lambda}$ and on $\Psi_{\Hat \Lambda}$.  For the 0-brane valise 
vector supermultiplet with field content vectors described by (\ref{V4}), the 
commutator algebra (\ref{V5}) is satisfied on $\Phi_{\Lambda}$ and on 
$\Psi_{\Hat \Lambda}$.

\section{A 0-brane-World Formulation of the Off-Shell SUSY Auxiliary Field Problem} 
$~~~$
Let $\Phi{}_{\Lambda}(\tau) $, and $\Psi{}_{\Hat {\Lambda}}(\tau)$ denote arbitrary
bosonic and fermionic sets of functions.  All the bosonic functions
satisfy the equation
\be
\Phi{}_{\Delta} (\tau_1) \, \Phi{}_{\Lambda} (\tau_2) ~=~ + 
 \Phi{}_{\Lambda} (\tau_2)  \, \Phi{}_{\Delta} (\tau_1)  ~~~,
\label{Bs0n}
\ee
and all the fermionic functions satisfy the equation
\be
\Psi{}_{{\Hat \Delta}} (\tau_1) \, \Psi{}_{\Hat \Lambda} (\tau_2) ~=~ - 
 \Psi{}_{\Hat \Lambda} (\tau_2)  \, \Psi{}_{\Hat \Delta} (\tau_1)  ~~~,
\label{Fm0n}
\ee
The off-shell auxiliary field problem then asks that one determine all
sets of bosonic functions $\Phi{}_{\Lambda}(\tau) $, sets of fermionic
functions $\Psi{}_{\Hat {\Lambda}}(\tau)$, constant coefficients $\left( 
{\rm L}{}_{\Lambda}\right){}_{a}{}^{\Hat {\Lambda}}$, and $\left( {\rm 
R}{}^{\Lambda} \right){}_{\Hat {\Lambda}}{\,}_{a}$ (where these coefficients 
are constructed from Lorentz invariant tensors and gamma matrices)
such that the equations 
\be \eqalign{
{\rm D}{}_a \, \Phi{}_{\Lambda}  ~=~ i\,  \left( {\rm L}{}_{\Lambda}\right){}_{a}
{}^{\Hat {\Lambda}} \, \, \Psi{}_{\Hat {\Lambda}}
~~~,~~~
{\rm D}{}_{a} \Psi{}_{\Hat {\Lambda}} ~=~  \left( {\rm R}{}^{\Lambda} \right)
{}_{\Hat {\Lambda}}{\,}_{a} \, \pa_{\tau} \, \Phi{}_{\Lambda}  ~~~,
} \label{covVAL2}
\ee 
necessarily implies 
\be
\{ \,  {\rm D}{}_{a} ~,~  {\rm D}{}_{b} \, \} ~=~ i \, 2 \, (\gamma^0){}_{a \, b}  
\pa_{\t}  ~~~,
\label{SUSYeq2}
\ee
and this should be done in an irreducible manner in the space of field
vectors.  With the exception of the 4D, $\cal N$ = 1 double tensor 
multiplet\footnote{See the work in \cite{G-1} for details.}, there is a solution 
for (\ref{Bs0n}) - (\ref{SUSYeq2}) in the case of every studied supermultiplet 
known to these authors.  The solution to this problem is generally {\em {not}} 
known for either $\cal N$-extended supersymmetry or supersymmetry in 
higher space time dimensions than four.  

The most prominent case showing such a failure is the 4D, $\cal 
N$ = 4 Maxwell Supermultiplet.  Here the field content vectors take the 
respective forms
\be\eqalign{ {~~~}
\Phi{}_{\Lambda} ~&=~    \left(\, A_m, A{}^{\cal I}, \, B{}^{\cal I} , \,{\rm d}, \, 
F{}^{\cal I}, \, G{}^{\cal I}  \,\right)   ~~,~~~
\Psi{}_{\Hat {\Lambda}} ~=~ \left( \, \lambda{}_a , \, \psi{}_a{}^{\cal I} \,  
\,\right)   ~~~, \cr
} \label{Zhiz}
\ee
and these are written appropriately for the realization of one of the 
four supersymmetries in an off-shell manner.  The indices $\cal I$, $
\cal J$, etc. here and in the following discussion take on three values.   
The 0-brane version of an invariant action is given by 
\cite{adnkN4SYM2,adnkN4SYM1}
\be\eqalign{
{\mathcal{L}}~=~ &\frac{1}{2}(\partial_{\tau}A^{\cal I})(\partial_{\tau}A^{\cal I}) 
+ \frac{1}{2}(\partial_{\tau}B^{\cal I})(\partial_{\tau}B^{\cal I})
+\frac{1}{2}(\pa_{\tau}F^{\cal I})(\pa_{\tau}F^{\cal I})+\frac{1}{2}(\pa_{\tau}G^{\cal I})
(\pa_{\tau}G^{\cal I})
\cr
& +\frac{1}{2} (\partial_{\tau}A_m)  (\partial_{\tau}A_m) 
+\frac{1}{2}({\pa_{\tau} \rm d}) ({\pa_{\tau} \rm d}) 
+i\frac{1}{2}(\gamma^0)^{ab}\psi_{a}^{\cal I}\partial_{\tau}\psi_{b
}^{\cal I} + i \frac{1}{2}(\gamma^0)^{
cd}\lambda_{c} \pa_{\tau}  \lambda_{d} ~~~.
}  \label{n4actv} \ee
The four supercharges can be represented
by  ${\rm D}{}_{a}$ and  ${\rm D}{}_{a}^{\cal I}$ where
\be \eqalign{
{\rm D}_a A^{\cal J} ~=~&~ \psi_a^{\cal J}  ~~~~~~~~~~~, ~~~
{\rm D}_a B^{\cal J} ~=~ ~i \, (\gamma^5){}_a{}^b \, \psi_b^{\cal J}  ~~~~~~, \cr
{\rm D}_a F^{\cal J} ~=~&~ (\gamma^0){}_a{}^b \,  \psi_b^{\cal J}   
~~~, ~~~
{\rm D}_a G^{\cal J} ~=~ ~i \,(\gamma^5\gamma^0){}_a{}^b \, 
\psi_b^{\cal J}  ~~~,  \cr
{\rm D}_a \psi_b^{\cal J} ~=~ &~i\, (\gamma^0){}_{a \,b}\, \left( 
\partial_{\tau} A^{\cal J} \right) ~-~  (\gamma^5\gamma^0){}_{a 
\,b} \, \left( \partial_{\tau} B^{\cal J} \right) ~\cr
&-~ ~i \, C_{a\, b} \, \left( \partial_\tau
\,F^{\cal J}  \right)~+~ ~ (\gamma^5){}_{ a \, b} \, \left( \partial_\tau G^{\cal 
J} \right) ~~~~~~\,~, \cr
{\rm D}_a A_m  ~=~& (\gamma_m)_a{}^b \lambda_b  ~~~, ~~~ {\rm D}_a {\rm d} 
 \, =\,  i (\gamma^5 \gamma^0)_a{}^b \, \lambda_b 
 ~~~~~~~~~~~~~~, \cr
{\rm D}_a \lambda_b  ~=~& - i( \gamma^0 
\gamma^m )_{ab} \,  \left( \,  \pa_{\tau} A_m  \, \right)~+~ (\gamma^5
)_{ab} \,  \left( \, \pa_{\tau} {\rm d} \, \right)
~~~, \cr 
{\rm D}_a^{\cal I} A^{\cal J} ~=~& \delta^{\cal {I \, J}}~ \l_a - \epsilon^{\cal {I 
\, J}}_{~~~{\cal K}} \,\psi_a^{\cal K}
~~~~~, ~~~~  \cr
{\rm D}_a^{\cal I} B^{\cal J} ~=~& ~i \, (\gamma^5){}_a{}^b \, \left[~\delta^{IJ}
~ \l_b + \epsilon^{\cal {I \, J}}_{~~~{\cal K}} \,\psi_b^{\cal K}~ \right]  ~~~, \cr
{\rm D}_a^{\cal I} F^{\cal J} ~=~& ~ (\gamma^0){}_a{}^b \,  
{\big [}~ \delta^{\cal {I \, J}}~ \l_b - \epsilon^{\cal {I \, J}}_{~~~{\cal K}} \,\psi_b^{\cal K}~
{\big ]}   ~~~, \cr
{\rm D}_a^{\cal I} G^{\cal J} ~=~& ~i \,(\gamma^5\gamma^0){}_a{}^b 
\,  {\big  [}~-\delta^{IJ}~ \l_b + \epsilon^{\cal {I \, J}}_{~~~{\cal K}} \,
\psi_b^{\cal K}~ {\big ]}  ~~~, \cr
{\rm D}_a^{\cal I} \psi_b^{\cal J} ~=~& \delta^{\cal {I \, J}} {\big [}~ i \,  (  
\gamma^0  \gamma^{m} ){}_a{}_b \, (\,  \partial_{\tau}  
\, A{}_{m}   \, ) ~+~  (\gamma^5){
}_{a \,b} \,  \left( \pa_{\tau}  {\rm d}  \right) ~ {\big ]}  \cr
&+~\epsilon^{\cal {I \, J}}_{~~~{\cal K}} \,{\big [}~ i\, (\gamma^0){}_{a \,b}\,  
\left( \partial_{\tau} A^{\cal K} \right) ~+~ (\gamma^5\gamma^0)
{}_{a \,b} \, \left( \partial_{\tau} B^{\cal K} \right) ~\cr
&~~~~~~~~~~~-~ i \, C_{a\, b} \, \left( \partial_\tau F^{\cal K} \right)  
~-~ (\gamma^5){}_{ a \, b}\, \left( \partial_\tau G^{\cal K} \right)
~ {\big ]}  ~~, \cr
{\rm D}_a^{\cal I} \, A{}_{m} ~=~ & -(\gamma_m){}_a {}^b \,  
\psi_b^{\cal I}  ~~~, ~~~~
{\rm D}_a^{\cal I} \, {\rm d} ~=~ i \, (\gamma^5\gamma^0){}_a {}^b \, 
\,   \psi_b^{\cal I}  ~~~, \cr
{\rm D}_a^{\cal I} \l_b ~=~ & ~i\, (\gamma^0){}_{a \,b}\,  \left( 
\partial_\tau A^{\cal I} \right)~-~ ~ (\gamma^5\gamma^0)
{}_{a \,b} \, \left( \partial_\tau B^{\cal I} \right) ~\cr
&-~i \, C_{a\, b} \, \left(  \partial_\tau F^{\cal I} \right)  ~-~ (\gamma^5)
{}_{ a \, b} \, \left( \partial_\tau G^{\cal I} \right)  ~~~,
} \label{VIspecifZZ}
\ee
The three supersymmetries generated by D${}_a^{\cal I}$ are
on-shell.  If they were off-shell, the $\cal N$ = 4
extended version of (\ref{SUSYeq2}) would read as
\be  \eqalign{
\{ \,  {\rm D}{}_{a} ~,~  {\rm D}{}_{b} \, \} ~&=~ i \, 2 \, (\gamma^0){}_{a \, b}  \pa_{\t}  
~~, ~~
\{ \,  {\rm D}{}_{a} ~,~  {\rm D}{}_{b}^{\cal I} \, \} ~=~ 0  ~~, ~~ \cr
\{ \,  {\rm D}{}_{a}^{\cal I} ~,~  {\rm D}{}_{b}^{\cal J} \, \} ~&=~ i \, 2 \, \delta{}^{{\cal I}
\, {\cal J}}
(\gamma^0){}_{a \, b}  \pa_{\t}  ~~~.
}
\label{SUSYeq3}
\ee
The explicit forms of the coefficients in (\ref{covVAL2}) appropriate for
this theory can now be read off from the equations in (\ref{VIspecifZZ})
then via direct calculation, it is found  \cite{adnkN4SYM1} that only the 
first equation in (\ref{SUSYeq3}) is satisfied by the field content in (\ref{Zhiz}).

The strongest interpretation of the Siegel-Ro\v cek No-Go Theorem to this
1D valise formulation would involve claiming there exists no possible
extension of the field content vectors in (\ref{Zhiz}) such that the equations in 
(\ref{SUSYeq3}) can be satisfied.  To reach this result, however, requires
an assumption about the form of additional terms that must be added
to (\ref{n4actv}) as in the original discussion.

\section{Within the World of Adinkra Network Valise Supermultiplets} 
$~~~$
During the course of our efforts since the work of \cite{GRana2,GRana3}, 
we have produced evidence suggesting there exists a
way to apply the old Macintosh Mantra of ``Think Different'' to the
problem stated in the previous chapter.  

This alternative approach starts from networks that precisely encode the same kinematic 
information as the 0-brane-world description of valise supermultiplets.  
The graphical representations of these networks were given the name of 
``adinkras'' \cite{adnk1}.  Two examples of these are shown below. 
$$
\vCent
{\setlength{\unitlength}{1mm}
\begin{picture}(-20,-140)
\put(-50,-4){\bf {{$\cal R\,=$} \# 1}}
\put(30,-4){\bf {{$\cal R\,=$} \# 2}}
\put(-74,-39){\includegraphics[width=2.6in]{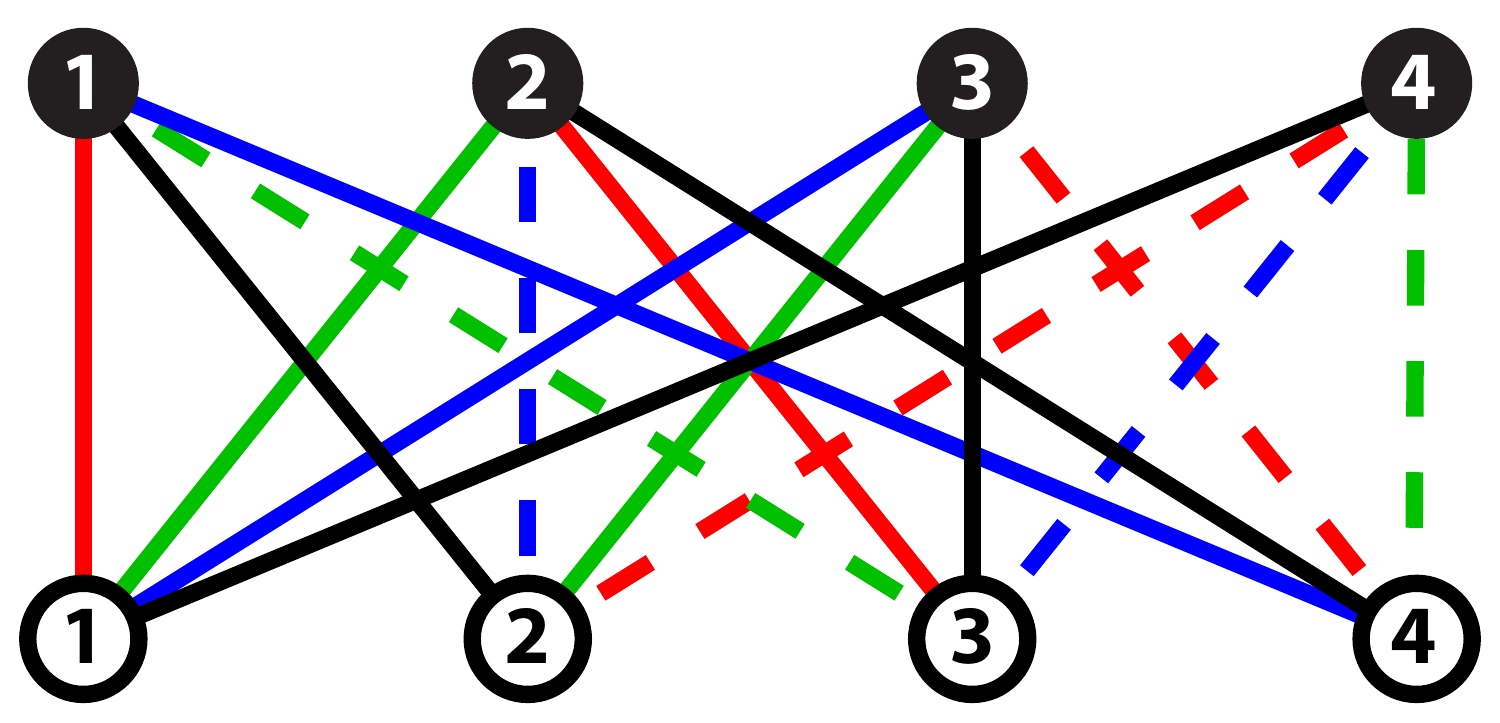}}
\put(6,-39.2){\includegraphics[width=2.6in]{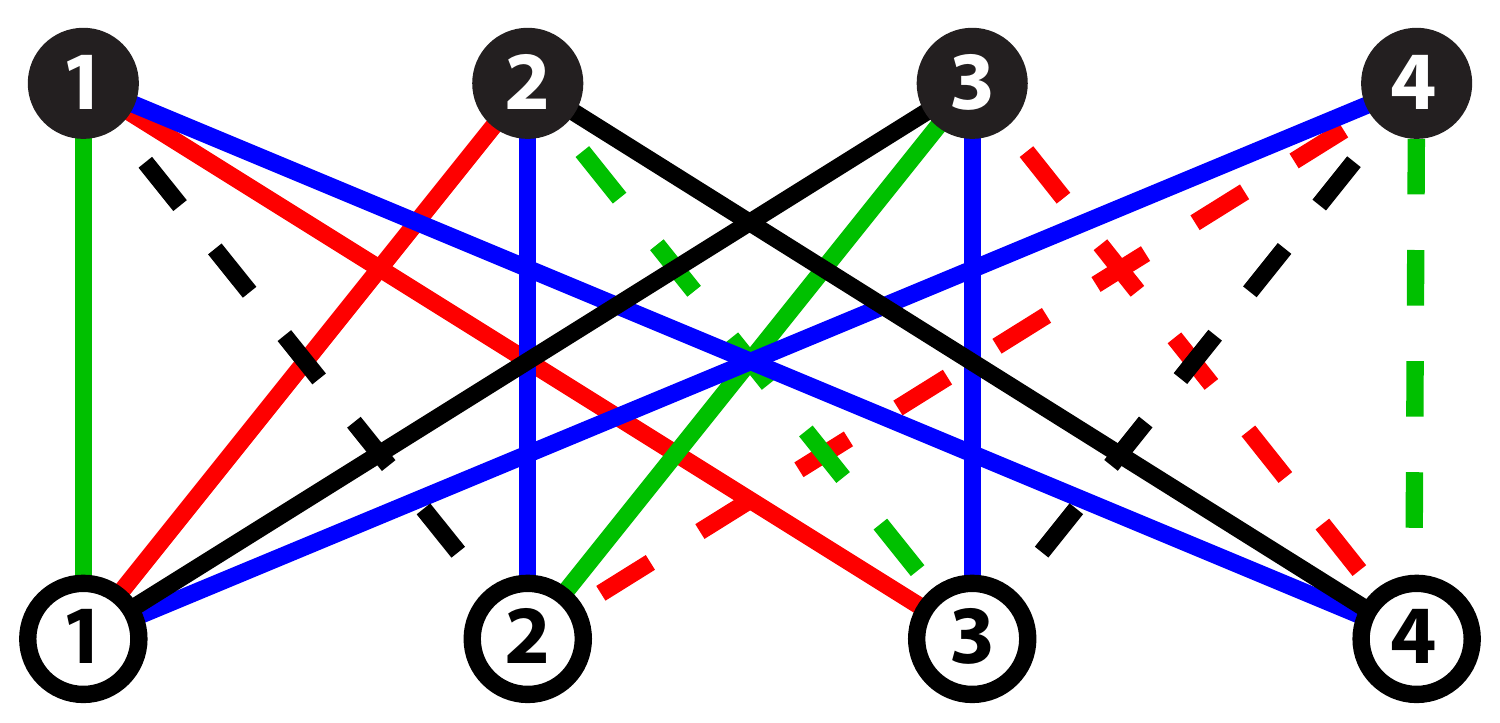}}
\put(-64,-47){\bf {Figure \# 1: Two valise adinkra graphs with node assignment}}
\end{picture}}
$$
\vskip1.8in 
 \noindent
We were led to these graphs by first discovering the adjacency matrices 
\cite{GRana2,GRana3} associated with them.  These adjacency matrices
satisfy a set of algebraic conditions we have named the ${ \bf{ {\cal GR} 
({\rm d}, \,  {\cal N})}}$ or ``Garden Algebra'' conditions and have been 
completely defined in the works of \cite{DFGHILM4,DFGHILM,DFGHILM2,DFGHILM3}.
Via a set of Feynman-like rules (see e.g. \cite{G-1}), these networks 
can be shown to be equivalent to equations
\be  \eqalign{
{\rm D}{}_{{}_{\rm I}} \Phi_i ~=~ i \, \left( {\rm L}{}_{{}_{\rm I}}\right) {}_{i \, 
{\hat k}}  \,  \Psi_{\hat k}
~~,~~
{\rm D}{}_{{}_{\rm I}} \Psi_{\hat k} ~=~  \left( {\rm R}{}_{{}_{\rm I}}\right)
{}_{{\hat k} \, i}  \, \pa_{\tau} \, \Phi_{i}  ~\implies~
\left\{ \, {\rm D}{}_{{}_{\rm I}}  ~,~ {\rm D}{}_{{}_{\rm I}}  \, \right\} 
~=~ i \, 2 \, \delta{}_{{}_{\rm I} \, {}_{\rm J}} \, \pa_{\tau} ~~.
} \label{chiD0EXX}
\ee

\subsection{Chiral Supermultiplet Adinkra Network Valise Off-Shell} 
$~~~$
In the case of the first adinkra network shown in (Fig.\ \#1), the L-matrices and
R-matrices take the forms given by
$$
\left( {\rm L}{}_{1}\right) {}_{i \, {\hat k}}   ~=~
\left[\begin{array}{cccc}
~1 & ~~0 &  ~~0  &  ~~0 \\
~0 & ~~0 &  ~~0  &  ~-\, 1 \\
~0 & ~~1 &  ~~0  &  ~~0 \\
~0 & ~~0 &  ~-\, 1  &  ~~0 \\
\end{array}\right] ~~~,~~~
\left( {\rm L}{}_{2}\right) {}_{i \, {\hat k}}   ~=~
\left[\begin{array}{cccc}
~0 & ~~1 &  ~~0  &  ~ \, \, 0 \\
~0 & ~~ 0 &  ~~1  &  ~~0 \\
-\, 1 & ~~ 0 &  ~~0  &  ~~0 \\
~ 0 & ~~~0 &  ~~0  &   -\, 1 \\
\end{array}\right]  ~~~,
$$
\be
\left( {\rm L}{}_{3}\right) {}_{i \, {\hat k}}   ~=~
\left[\begin{array}{cccc}
~0 & ~~0 &  ~~1  &  ~~0 \\
~0 & ~- \, 1 &  ~~0  &  ~~0 \\
~0 & ~~0 &  ~~0  &  -\, 1 \\
~1 & ~~0 &  ~~0  &  ~~0 \\
\end{array}\right] ~~~,~~~
\left( {\rm L}{}_{4}\right) {}_{i \, {\hat k}}   ~=~
\left[\begin{array}{cccc}
~0 & ~~0 &  ~~0  &  ~ \, \, 1 \\
~1 & ~~ 0 &  ~~0  &  ~~0 \\
~0 & ~~ 0 &  ~~1  &  ~~0 \\
~ 0 & ~~~1 &  ~~0  &   ~~0  \\
\end{array}\right]  ~~.
 \label{chiD0F}
\ee

$$
\left( {\rm R}{}_{1}\right) {}_{{\hat k} \, i}    ~=~
\left[\begin{array}{cccc}
~1 & ~~0 &  ~~0  &  ~~0 \\
~0 & ~~0 &  ~~1  &  ~~ 0 \\
~0 & ~~0 &  ~ 0  &  ~- 1 \\
~0 & -1 &  ~ 0  &  ~~0 \\
\end{array}\right] ~~~,~~~
\left( {\rm R}{}_{2}\right) {}_{{\hat k} \, i}    ~=~
\left[\begin{array}{cccc}
0 & ~~0 &  ~-\, 1  &  ~ \, \, 0 \\
~1 & ~~ 0 &  ~~0  &  ~~0 \\
~0 & ~~ 1 &  ~~0  &  ~~0 \\
~ 0 & ~~~0 &  ~~0  &   -\, 1 \\
\end{array}\right]  ~~~,
$$
\be
\left( {\rm R}{}_{3}\right) {}_{{\hat k} \, i}    ~=~
\left[\begin{array}{cccc}
~0 & ~~0 &  ~~0  &  ~1 \\
~0 & - \, 1 &  ~~0  &  ~0 \\
~1 & ~~0 &  ~~0  &  ~0 \\
~0 & ~~0 &   - 1  &  ~0 \\
\end{array}\right] ~~~,~~~
\left( {\rm R}{}_{4}\right) {}_{{\hat k} \, i}    ~=~
\left[\begin{array}{cccc}
~0 & ~~1 &  ~~0  &  ~ \, \, 0 \\
~0 & ~~ 0 &  ~~0  &  ~~1 \\
~0 & ~~ 0 &  ~~1  &  ~~0 \\
~ 1 & ~~~0 &  ~~0  &   ~~0  \\
\end{array}\right]  ~~~.
 \label{chiD0K}
\ee
These satisfy the Garden Algebra relationships
\be \eqalign{
{~~~~~}
 (\,{\rm L}_\rI\,)_i{}^\hj\>(\,{\rm R}_\rJ\,)_\hj{}^k + (\,{\rm L}_\rJ\,)_i{}^\hj\>(\,{\rm 
 R}_\rI\,)_\hj{}^k &= 2\,\delta_{\rI\rJ}\, \delta_i{}^k  ~~~~~,\cr
(\,{\rm R}_\rJ\,)_\hi{}^j\>(\,{\rm L}_\rI\,)_j{}^\hk + (\,{\rm R}_\rI\,
)_\hi{}^j\>(\,{\rm L}_\rJ\,)_j{}^\hk
&=~ 2\,  \delta{}_{\rI\rJ} \, \delta_\hi{}^\hk ~~.
}  \label{GDNAlge1}
\ee

\subsection{Chiral Supermultiplet Adinkra Network Valise On-Shell} 
$~~~$
If we delete the open nodes denoted by 3 and 4 in
the first adinkra labelled as $\cal R\,=$ \# 1 as well as eliminate
all the links associated with those nodes we find,
$$
\left( {\rm L}{}_{1}\right) {}_{i \, {\hat k}}   ~=~
\left[\begin{array}{cccc}
~1 & ~~0 &  ~~0  &  ~~0 \\
~0 & ~~0 &  ~~0  &  ~-\, 1 \\
\end{array}\right] ~~~,~~~
\left( {\rm L}{}_{2}\right) {}_{i \, {\hat k}}   ~=~
\left[\begin{array}{cccc}
~0 & ~~1 &  ~~0  &  ~ \, \, 0 \\
~0 & ~~ 0 &  ~~1  &  ~~0 \\
\end{array}\right]  ~~~,
$$
\be
\left( {\rm L}{}_{3}\right) {}_{i \, {\hat k}}   ~=~
\left[\begin{array}{cccc}
~0 & ~~0 &  ~~1  &  ~~0 \\
~0 & ~- \, 1 &  ~~0  &  ~~0 \\
\end{array}\right] ~~~,~~~
\left( {\rm L}{}_{4}\right) {}_{i \, {\hat k}}   ~=~
\left[\begin{array}{cccc}
~0 & ~~0 &  ~~0  &  ~ \, \, 1 \\
~1 & ~~ 0 &  ~~0  &  ~~0 \\
\end{array}\right]  ~~~,
 \label{chiD0N}
\ee
$$
\left( {\rm R}{}_{1}\right) {}_{{\hat k} \, i}   ~=~
\left[\begin{array}{cc}
~1 & ~~0  \\
~0 & ~~0  \\
~0 & ~~0  \\
~0 & -1  \\
\end{array}\right] ~~~,~~~
\left( {\rm R}{}_{2}\right) {}_{{\hat k} \, i}   ~=~
\left[\begin{array}{cc}
0 & ~~0  \\
~1 & ~~ 0  \\
~0 & ~~ 1  \\
~ 0 & ~~~0  \\
\end{array}\right]  ~~~,
$$
\be
\left( {\rm R}{}_{3}\right) {}_{{\hat k} \, i}   ~=~
\left[\begin{array}{cc}
~0 & ~~0  \\
~0 & - \, 1  \\
~1 & ~~0  \\
~0 & ~~0  \\
\end{array}\right] ~~~,~~~
\left( {\rm R}{}_{4}\right) {}_{{\hat k} \, i}   ~=~
\left[\begin{array}{cc}
~0 & ~~1  \\
~0 & ~~ 0  \\
~0 & ~~ 0  \\
~ 1 & ~~~0   \\
\end{array}\right]  ~~.
 \label{chiD0O}
\ee
Given the matrices in (\ref{chiD0N}) and (\ref{chiD0O}) we find the following relations hold
\be \eqalign{
{~~~~~}
 (\,{\rm L}_\rI\,)_i{}^\hj\>(\,{\rm R}_\rJ\,)_\hj{}^k + (\,{\rm L}_\rJ\,)_i{}^\hj\>(\,{\rm 
 R}_\rI\,)_\hj{}^k &= 2\,\delta_{\rI\rJ}\, \delta_i{}^k  ~~,\cr
(\,{\rm R}_\rJ\,)_\hi{}^j\>(\,{\rm L}_\rI\,)_j{}^\hk + (\,{\rm R}_\rI\,
)_\hi{}^j\>(\,{\rm L}_\rJ\,)_j{}^\hk
&=~  \delta{}_{\rI\rJ} \, \delta_\hi{}^\hk 
 ~+~  \ [\, {\vec {\a}} {\b}^1 \, ]_{\rI\rJ}\,  \cdot \,
(\, {\vec {\a}} {\b}^1 \, )_\hi{}^\hk   
~~.
}  \label{NotGDNAlg1}
\ee
The six 4 $\times$ 4 matrices ${\vec {\a}}$ and ${\vec {\b}}$ were defined in
the work of \cite{G-1}.

It can be seen that the matrices of (\ref{chiD0F}) and (\ref{chiD0K}) are d $\times$ d
matrices with d = 4.  On the other hand the matrices in (\ref{chiD0N}) are d${
}_L$ $\times$ d${}_R$, and the matrices in (\ref{chiD0O}) are d${}_R$ $\times$ 
$d{}_L$ where d${}_L$ = 2 and d${}_R$ = 4.  We refer to matrices of this sort of 
structure as representatives of the ${ \bf{ {\cal GR} ({\rm d}_L, \, {\rm d}_R, \,  {\cal 
N})}}$ algebra.

\subsection{Vector Supermultiplet Adinkra Network Valise Off-Shell} 
$~~~$
In the case of the second adinkra network shown in (Fig.\ \#1), the L-matrices 
and R-matrices take the forms 
$$
\left( {\rm L}{}_{1}\right) {}_{i \, {\hat k}}   ~=~
\left[\begin{array}{cccc}
~0 & ~1 &  ~ 0  &  ~ 0 \\
~0 & ~0 &  ~0  &  -\,1 \\
~1 & ~0 &  ~ 0  &  ~0 \\
~0 & ~0 &  -\, 1  &  ~0 \\
\end{array}\right] ~~~,~~~
\left( {\rm L}{}_{2}\right) {}_{i \, {\hat k}}   ~=~
\left[\begin{array}{cccc}
~1 & ~ 0 &  ~0  &  ~ 0 \\
~0 & ~ 0 &  ~1  &  ~ 0 \\
 ~0 & - \, 1 &  ~0  &   ~ 0 \\
~0 & ~0 &  ~0  &  -\, 1 \\
\end{array}\right]  ~~~,
$$
\be {~~~~}
\left( {\rm L}{}_{3}\right) {}_{i \, {\hat k}}   ~=~
\left[\begin{array}{cccc}
~0 & ~0 &  ~ 0  &  ~ 1 \\
~0 & ~1 &  ~0  &   ~0 \\
~0 & ~0 &  ~ 1  &  ~0 \\
~1 & ~0 &  ~0  &  ~0 \\
\end{array}\right] ~~~~~~,~~~
\left( {\rm L}{}_{4}\right) {}_{i \, {\hat k}}   ~=~
\left[\begin{array}{cccc}
~0 & ~0 &  ~1  &  ~ 0 \\
-\,1 & ~ 0 &  ~0  &  ~ 0 \\
 ~0 & ~0 &  ~0  &   - \, 1 \\
~0 & ~1 &  ~0  &  ~  0 \\
\end{array}\right]  ~~~,
\label{V1D0E}
\ee
$$
\left( {\rm R}{}_{1}\right) {}_{{\hat k} \, i}   ~=~
\left[\begin{array}{cccc}
~0 & ~~0 &  ~1  &  ~0 \\
~1 & ~~0 &  ~0  &  ~ 0 \\
~0 & ~~0 &  ~ 0  & ~- 1 \\
~0 & -1 &  ~ 0  &  ~0 \\
\end{array}\right] ~~~,~~~
\left( {\rm R}{}_{2}\right) {}_{{\hat k} \, i}   ~=~
\left[\begin{array}{cccc}
1 & ~0 &  ~0  &  ~ \, \, 0 \\
~0 & ~ 0 &  -\,1  &  ~~0 \\
~0 & ~ 1 &  ~~0  &  ~~0 \\
~ 0 & ~0 &  ~~0  &   -\, 1 \\
\end{array}\right]  ~~~,
$$
\be \eqalign{ {~~~}
\left( {\rm R}{}_{3}\right) {}_{{\hat k} \, i}   ~=~
\left[\begin{array}{cccc}
~0 & ~~0 &  ~~0  &  ~1 \\
~0 &   ~1 &  ~~0  &  ~0 \\
~0 & ~~0 &  ~~1  &  ~0 \\
~1 & ~~0 &   ~~0  &  ~0 \\
\end{array}\right] ~~~,~~~
\left( {\rm R}{}_{4}\right) {}_{{\hat k} \, i}   ~=~
\left[\begin{array}{cccc}
~0 & -1 &  ~0  &  ~ \, \, 0 \\
~0 & ~\, 0 &  ~0  &  ~~1 \\
~1 & ~\, 0 &  ~0  &  ~~0 \\
~0 & ~\, 0 &  - 1  &   ~~0  \\
\end{array}\right]  ~~~.
}   \label{V1D0HO} 
\ee
These also satisfy the relationships
\be \eqalign{
{~~~~~}
 (\,{\rm L}_\rI\,)_i{}^\hj\>(\,{\rm R}_\rJ\,)_\hj{}^k + (\,{\rm L}_\rJ\,)_i{}^\hj\>(\,{\rm 
 R}_\rI\,)_\hj{}^k &= 2\,\delta_{\rI\rJ}\, \delta_i{}^k  ~~~~~,\cr
(\,{\rm R}_\rJ\,)_\hi{}^j\>(\,{\rm L}_\rI\,)_j{}^\hk + (\,{\rm R}_\rI\,
)_\hi{}^j\>(\,{\rm L}_\rJ\,)_j{}^\hk
&=~ 2\,  \delta{}_{\rI\rJ} \, \delta_\hi{}^\hk ~~.
}  \label{GDNAlge2}
\ee

\subsection{Vector Supermultiplet Adinkra Network Valise On-Shell} 
$~~~$
If we erase the fourth open node and its associated links,
the forms of the associated adjacency-like matrices become
$$
\left( {\rm L}{}_{1}\right) {}_{i \, {\hat k}}   ~=~
\left[\begin{array}{cccc}
~0 & ~1 &  ~ 0  &  ~ 0 \\
~0 & ~0 &  ~0  &  -\,1 \\
~1 & ~0 &  ~ 0  &  ~0 \\
\end{array}\right] ~~~,~~~
\left( {\rm L}{}_{2}\right) {}_{i \, {\hat k}}   ~=~
\left[\begin{array}{cccc}
~1 & ~ 0 &  ~0  &  ~ 0 \\
~0 & ~ 0 &  ~1  &  ~ 0 \\
 ~0 & - \, 1 &  ~0  &   ~ 0 \\
\end{array}\right]  ~~~,
$$
\be {~~~~}
\left( {\rm L}{}_{3}\right) {}_{i \, {\hat k}}   ~=~
\left[\begin{array}{cccc}
~0 & ~0 &  ~ 0  &  ~ 1 \\
~0 & ~1 &  ~0  &   ~0 \\
~0 & ~0 &  ~ 1  &  ~0 \\
\end{array}\right] ~~~~~~,~~~
\left( {\rm L}{}_{4}\right) {}_{i \, {\hat k}}   ~=~
\left[\begin{array}{cccc}
~0 & ~0 &  ~1  &  ~ 0 \\
-\,1 & ~ 0 &  ~0  &  ~ 0 \\
 ~0 & ~0 &  ~0  &   - \, 1 \\
\end{array}\right]  ~~~,
\label{V1D0E2}
\ee

$$
\left( {\rm R}{}_{1}\right) {}_{ {\hat k} \, i}   ~=~
\left[\begin{array}{ccc}
~0 & ~0 &  ~1   \\
~1 & ~ 0 &  ~0   \\
 ~0 & ~0 &  ~0   \\
 ~0 & -1 &  ~0   \\
\end{array}\right] ~~~~~~,~~~
\left( {\rm R}{}_{2}\right) {}_{ {\hat k} \, i}   ~=~
\left[\begin{array}{ccc}
~1 & ~0 &  ~0   \\
~0 & ~ 0 &  -1   \\
 ~0 & ~1 &  ~0   \\
 ~0 & ~ 0 &  ~0   \\
\end{array}\right]  ~~~,
$$

\be
\left( {\rm R}{}_{3}\right) {}_{ {\hat k} \, i}   ~=~
\left[\begin{array}{ccc}
~0 & ~0 &  ~0   \\
~0 & ~ 1 &  ~0   \\
 ~0 & ~0 &  ~1   \\
 ~1 & ~ 0 &  ~0   \\
\end{array}\right] ~~~~~~,~~~
\left( {\rm R}{}_{4}\right) {}_{ {\hat k} \, i}   ~=~
\left[\begin{array}{ccc}
~0 & -\,1 &  ~1   \\
~0 & ~ 0 &  ~0   \\
 ~1 & ~0 &  ~0   \\
 ~0 & ~ 0 &  -\,1   \\
\end{array}\right]  ~~~,
\label{V1D0HOz}   
\ee
The matrices in (\ref{V1D0E}) and (\ref{V1D0HO}) lead to the following 
relations
\be \eqalign{
{~~~~~}
 (\,{\rm L}_\rI\,)_i{}^\hj\>(\,{\rm R}_\rJ\,)_\hj{}^k + (\,{\rm L}_\rJ\,)_i{}^\hj\>(\,{\rm 
 R}_\rI\,)_\hj{}^k &= 2\,\delta_{\rI\rJ}\, \delta_i{}^k  ~~,\cr
(\,{\rm R}_\rJ\,)_\hi{}^j\>(\,{\rm L}_\rI\,)_j{}^\hk + (\,{\rm R}_\rI\,)_\hi{}^j\>(\,{\rm 
L}_\rJ\,)_j{}^\hk &=~ \fracm 32 \, {\delta}_{\rI\rJ}\,  (\,  {\bf I}_4  \,)_\hi{}^\hk 
~-~ \fracm 12 \,  [\,  {\vec \a} \,  \b^2  \, ]_{\rI\rJ}\, \cdot \,  (\,  {\vec \a} \,  \b^2 
\, )_\hi{}^\hk  \cr
&~~~~+~ \fracm 12 \,  [\,   {\vec \a} \,  \b^1 \, ]_{\rI\rJ}\, \cdot \,  (\,  {\vec \a} \, 
\b^1   \, )_\hi{}^\hk  \cr
&~~~~+~ \fracm 12 \,  [\,   {\vec \a} \,  \b^3 \, ]_{\rI\rJ}\, \cdot \,  (\,  {\vec \a} \, 
\b^3   \, )_\hi{}^\hk  ~~.
}  \label{NotGDNAlg4}
 \ee

At this stage, it is obvious that there are some interesting correlations
between the calculations done from the $\gamma$-matrices of a 0-brane-world 
starting point and similar calculations done from the basis of the adjacency
matrices of an adinkra network-world starting point.

For the adinkra network valise chiral supermultiplet with adjacency matrices 
described by (\ref{chiD0F}), and (\ref{chiD0K}) the commutator algebra 
shown in (\ref{chiD0EXX}) is satisfied on $\Phi_i$ and on $ \Psi_{\hat k}$.  
For the adinkra network valise chiral supermultiplet with adjacency matrices 
described by (\ref{chiD0N}) and (\ref{chiD0N}) the commutator algebra 
shown in (\ref{chiD0EXX}) is satisfied on $\Phi_i$, but {\em {not}} on $\Psi_{
\hat k}$ due to the second line of (\ref{NotGDNAlg1}).

For the adinkra network valise vector supermultiplet with adjacency matrices 
described by (\ref{V1D0E}) and (\ref{V1D0HO}), the commutator algebra 
shown in (\ref{chiD0EXX}) is satisfied on $\Phi_i$ and on $ \Psi_{\hat k}$.  
For the adinkra network valise chiral supermultiplet with adjacency matrices 
described by (\ref{V1D0E2}) and (\ref{V1D0HOz}) the commutator algebra 
shown in (\ref{chiD0EXX}) is satisfied on $\Phi_i$, but {\em {not}} on $\Psi_{
\hat k}$ due to the second line of (\ref{NotGDNAlg4}).

\section{An Adinkra Network-World Formulation of the
Off-Shell SUSY Auxiliary Field Problem} 
$~~~$
The off-shell problem in the world of 0-brane valise supermultiplets can 
be recast into an equivalent one involving adinkra valise networks.  There 
is one important difference however.   As the starting point is in terms of 
adinkra networks, there is no information a priori about Lorentz representations.

Let $\Phi{}_{i}(\tau) $, and $\Psi{}_{\hat i}(\tau)$ denote arbitrary
bosonic and fermionic sets of functions associated with the nodes of
a valise adinkra.  All the bosonic functions satisfy the equation
\be
\Phi{}_{i} (\tau_1) \, \Phi{}_{j} (\tau_2) ~=~ + 
 \Phi{}_{j} (\tau_2)  \, \Phi{}_{i} (\tau_1)  ~~~,
\label{Bs0nadnk}
\ee
and all the fermionic functions satisfy the equation
\be
\Psi{}_{{\hat i}} (\tau_1) \, \Psi{}_{\hat k} (\tau_2) ~=~ - 
 \Psi{}_{\hat k} (\tau_2)  \, \Psi{}_{\hat i} (\tau_1)  ~~~.
\label{Fm0nadnk}
\ee
The off-shell auxiliary field problem then asks that one determine all
sets of bosonic functions $\Phi{}_{i}(\tau) $, sets of fermionic functions 
$\Psi{}_{\Hat k}(\tau)$, associated matrices $\left( {\rm L}{
}_{{}_{\rm I}}\right) {}_{i \, {\hat k}} $ and $ \left( {\rm R}{}_{{}_{\rm I}}
\right) {}_{{\hat k} \, i} $ to be used in the equations
\be  \eqalign{
{\rm D}{}_{{}_{\rm I}} \Phi_i ~=~ i \, \left( {\rm L}{}_{{}_{\rm I}}\right) {}_{
i \, {\hat k}}  \,  \Psi_{\hat k}
~~,~~
{\rm D}{}_{{}_{\rm I}} \Psi_{\hat k} ~=~  \left( {\rm R}{}_{{}_{\rm I}}\right) 
{}_{{\hat k} \, i}  \, \pa_{\tau} \, \Phi_{i}  ~~.
} \label{chiD0Ew}
\ee
Since the definitions of these L-matrices and R-matrices rely
on the adinkras networks we have
\be \eqalign{
 (\,{\rm L}_\rI\,)_i{}^\hj\>(\,{\rm R}_\rJ\,)_\hj{}^k + (\,{\rm L}_\rJ\,)_i{}^\hj\>(\,{\rm 
 R}_\rI\,)_\hj{}^k &= 2\,\d_{\rI\rJ}\,\d_i{}^k~~,\cr
 (\,{\rm R}_\rJ\,)_\hi{}^j\>(\, {\rm L}_\rI\,)_j{}^\hk + (\,{\rm R}_\rI\,)_\hi{}^j\>(\,{\rm 
 L}_\rJ\,)_j{}^\hk
  &= 2\,\d_{\rI\rJ}\,\d_\hi{}^\hk~~,
}  \label{GarDNAlg2}
 \ee
which then imply the  result
\be
\left\{ \, {\rm D}{}_{{}_{\rm I}}  ~,~ {\rm D}{}_{{}_{\rm I}}  \, \right\} 
~=~ i \, 2 \, {}_{{}_{\rm I} \, {}_{\rm J}} \, \pa_{\tau} ~~,
\label{chiD0EZ}
\ee
on both bosonic and fermionic field variables.  According to our previous 
studies of adinkras, this is a solved problem.  

Thus the question becomes, ``How can this information be used to address 
the off-shell problem in adinkra network world?''  Stated another way, if one 
is solely given the information in (\ref{chiD0N}) and (\ref{chiD0O}) how does 
one recover (\ref{chiD0F}) and (\ref{chiD0K}) for the chiral supermultiplet 
adinkra?  Alternately, given solely the information in (\ref{V1D0E2}) and
(\ref{V1D0HOz}) how does one recover (\ref{V1D0E}) and (\ref{V1D0HO})
for the vector supermultiplet adinkra?

This will be addressed in the next chapter with the introduction of the
concept of ``On-Shell Adinkra-Network Deformations.''

\section{On-Shell Adinkra Network Deformations} 
$~~~$
When one reviews the arguments and equations of chapter four in comparison
with those in chapter six, it may seems as though the problems are the same.

The 0-brane-world formulation begins with bosonic variables $ \Phi{}_{\Lambda}$ 
and fermionic variables $ \Psi{}_{\Hat \Lambda}$ in equations of the form
\be \eqalign{
{\rm D}{}_a \, \Phi{}_{\Lambda}  ~=~ i\,  \left( {\rm L}{}_{\Lambda}\right){}_{a}
{}^{\Hat {\Lambda}} \, \, \Psi{}_{\Hat {\Lambda}}
~~~,~~~
{\rm D}{}_{a} \Psi{}_{\Hat {\Lambda}} ~=~  \left( {\rm R}{}^{\Lambda} \right)
{}_{\Hat {\Lambda}}{\,}_{a} \, \pa_{\tau} \, \Phi{}_{\Lambda}  ~~~,
} \label{covVAL2p}
\ee 
that ought then necessarily imply
\be
\{ \,  {\rm D}{}_{a} ~,~  {\rm D}{}_{b} \, \} ~=~ i \, 2 \, (\gamma^0){}_{a \, b}  
\pa_{\t}  ~~~,
\label{SUSYeq2p}
\ee
to describe an off-shell supermultiplet.

The adinkra-network world formulation begins with bosonic variables $\Phi_i $ 
and fermionic variables $ \Psi_{\hat k}$  in equations of the form
\be  \eqalign{
{\rm D}{}_{{}_{\rm I}} \Phi_i ~=~ i \, \left( {\rm L}{}_{{}_{\rm I}}\right) {}_{
i \, {\hat k}}  \,  \Psi_{\hat k}
~~,~~
{\rm D}{}_{{}_{\rm I}} \Psi_{\hat k} ~=~  \left( {\rm R}{}_{{}_{\rm I}}\right) 
{}_{{\hat k} \, i}  \, \pa_{\tau} \, \Phi_{i}  ~~,
} \label{chiD0Ewp}
\ee
that ought then necessarily imply
\be
\left\{ \, {\rm D}{}_{{}_{\rm I}}  ~,~ {\rm D}{}_{{}_{\rm I}}  \, \right\} 
~=~ i \, 2 \, \delta{}_{{}_{\rm I} \, {}_{\rm J}} \, \pa_{\tau} ~~.
\label{chiD0EZp}
\ee
The similarities between (\ref{covVAL2p}) and (\ref{SUSYeq2p}) on the 
one hand and (\ref{chiD0Ewp}) and (\ref{chiD0EZp}) on the other are 
striking.  However, computationally and operationally there are subtle 
differences.  

In order to go from (\ref{covVAL2p}) to (\ref{SUSYeq2p}) one must
\newline \indent
(1a.) make ansatz\" e for the coefficients $ \left( {\rm L}{}_{\Lambda}\right)$
and $ \left( {\rm R}{}^{\Lambda} \right)$,
\newline \indent
(1b.) make ansatz\" e for the field content vectors $\Phi{}_{\Lambda} $ and $\Psi{}_{\Hat {\Lambda}} $,
\newline \indent
(1c.) calculate a set of matrix equations involving $ \left( {\rm L}{}_{\Lambda}\right)$
and $ \left( {\rm R}{}^{\Lambda} \right)$ \newline \indent
$~~~~~~~$ to evaluate on all the bosonic fields, and
\newline \indent
(1d.) calculate a set of Fierz identities involving $ \left( {\rm L}{}_{\Lambda}\right)$
and $ \left( {\rm R}{}^{\Lambda} \right)$ \newline \indent
$~~~~~~~$ to evaluate on all the fermionic fields.
\newline
This last step is so because the quantities 
$ \left( {\rm L}{}_{\Lambda}\right)$ and $\left( {\rm R}{}^{\Lambda} \right)$ 
are constructed from $\gamma$-matrices and the evaluation of  
(\ref{SUSYeq2p}) acting on fermions in a valise supermultiplet requires 
evaluation of Fierz Identities.  

In order to go from (\ref{chiD0Ewp}) to (\ref{chiD0EZp}) one must
\newline \indent
(2a.) make ansatz\" e for the coefficients $\left( {\rm L}{}_{{}_{\rm I}}\right) $
and $ \left( {\rm R}{}_{{}_{\rm I}}\right)$,
\newline \indent
(2b.) make ansatz\" e for the field content vectors $\Phi_i$ and $\Psi_{\hat k}$,
\newline \indent
(2c.) calculate a set of matrix equations involving  $\left( {\rm L}{}_{{}_{\rm I}}\right) $
and $ \left( {\rm R}{}_{{}_{\rm I}}\right)$
 \newline \indent
$~~~~~~~$ to evaluate on all the bosonic fields, and
\newline \indent
(2d.) calculate a set of matrix equations involving  $\left( {\rm L}{}_{{}_{\rm I}}\right) $
and $ \left( {\rm R}{}_{{}_{\rm I}}\right)$
 \newline \indent
$~~~~~~~$ to evaluate on all the fermionic fields.
\newline
Obtaining (\ref{chiD0EZp}) from 
(\ref{chiD0Ewp}) does {\em {not}} require knowledge of Fierz Identities
as the quantities $ \left( {\rm L}{}_{{}_{\rm I}}\right)$ and $ \left( {\rm R}{
}_{{}_{\rm I}}\right)$ are constructed from adinkra network related
matrices and only matrix multiplication is required to
evaluate (\ref{chiD0EZp}) on both bosons and fermions.  This latter
distinction makes for a substantial difference in the design of algorithms
to search for possible auxiliary fields.

Another savings in required computation occurs because of differences
in field content vectors required for their respective ansat\" e.  In the case
of $\Phi{}_{\Lambda} $ and $\Psi{}_{\Hat {\Lambda}} $ one must include
data about the space-time spin of the component fields.   In the case of 
$\Phi_i$ and $\Psi_{\hat k}$ all one has to do is to require that the range 
of their respective indices goes from 1 to multiples of 4.  As the spin
bundle information of the fields is considerable, any calculation involving
them must keep track of this information.

One of the results of our previous work is it appears
such spin-bundle information emerges from the adinkra
networks.  In other words, by working with component fields that only 
depend on time and possess an SU(2) $\times$SU(2) symmetry, 
embedded within four color networks seems to allow the isospin of the
network to completely carry the spin bundle information for free. 

We consider how to create algorithms to go from an on-shell
adinkra network to an off-shell one.

As we have seen (see (\ref{chiD0N}) and (\ref{chiD0O}) for the 
on-shell chiral adinkra network and  (\ref{V1D0E}) and  (\ref{V1D0HO})
for the on-shell vector adinkra network), in on-shell adinkra networks
only some of the rows or columns are given due to the on-shell nature
of the representation.  So the unknown entries in the L-matrices and
R-matrices can be represented by real parameters we will denote
by the symbol $\ell$.  These may be used to augment the rows and
columns of the L-matrices and R-matrices until one reaches a $4p$
$\times$ $4p$ matrix in all cases for some integer $p$.  This is
explicitly shown in equations (\ref{MtrXL}), (\ref{MtrXL2}), (\ref{V1D0Y1}),
and (\ref{V1D0Y2}) below.  The problem of going from the on-shell
adinkra network to a corresponding off-shell one, has now been
reduced to the problem of determining the values of the $\ell$-parameters
in the augmented matrices so as to satisfy the conditions in (\ref{GDNAlge1}).

It is amusing to note that this problem is roughly analogous to a cryptographic
one.  The on-shell forms of the L-matrices and R-matrices all together for 
any particular valise supermultiplet play the role of an encrypted message 
and finding the corresponding off-shell L-matrices and R-matrices is analogous 
to decoding the message.  

We now need to specify a series of operations to achieve this.  The key to 
achieving this is the Garden Algebra (\ref{GDNAlge1}).  
These conditions can be separated into four different parts
\be \eqalign{
{~~~~~}
 (\,{\rm L}_\rI\,)_i{}^\hj\>(\,{\rm R}_\rJ\,)_\hj{}^k + (\,{\rm L}_\rJ\,)_i{}^\hj\>(\,{\rm 
 R}_\rI\,)_\hj{}^k &= \,0 ~~{\rm {where}} ~ I ~\ne ~ J ~~~.
}  \label{GDNAlge1SPLT1}
\ee
\be \eqalign{
{~~~~~}
(\,{\rm R}_\rJ\,)_\hi{}^j\>(\,{\rm L}_\rI\,)_j{}^\hk + (\,{\rm R}_\rI\,
)_\hi{}^j\>(\,{\rm L}_\rJ\,)_j{}^\hk &=  \,0 ~~{\rm {where}} ~ I ~\ne ~ J ~~~.
}  \label{GDNAlge1SPLT2}
\ee
\be \eqalign{
{~~~~~}
 (\,{\rm L}_\rI\,)_i{}^\hj\>(\,{\rm R}_\rJ\,)_\hj{}^k &=~ \delta_i{}^k  
  ~~{\rm {where}} ~ I ~= ~ J ~~~.
}  \label{GDNAlge1SPLT3}
\ee
\be \eqalign{
{~~~~~}
(\,{\rm R}_\rJ\,)_\hi{}^j\>(\,{\rm L}_\rI\,)_j{}^\hk  &=~ \delta_\hi{}^\hk 
  ~~{\rm {where}} ~ I ~= ~ J ~~~.
}  \label{GDNAlge1SPLT4}
\ee
In the subsequent discussion, we show how the use of these for the
augmented on-shell L-matrices and R-matrices leads from on-shell
results to off-shell ones in the case of the chiral and vector adinkra
networks.

\subsection{On-Shell Chiral Valise Matrix Deformation} 
$~~~$
Define four matrices ${\rm L}{}_{{}_{\rm I}} $ ${\rm {where ~ I ~ = 1, \, 2, \, 3, \, ~or ~4}} $
that also depend on eight continuous real variables denoted by $\ell_{3 \, 1}$,
$\ell_{3 \, 2}$, $\ell_{3 \, 3}$, $\ell_{3 \, 4}$, $\ell_{4 \, 1}$, $\ell_{4 \, 2}$,
$\ell_{4 \, 3}$, and $\ell_{4 \, 4}$,
 via the four equations
\be   \eqalign{
\left( {\rm L}{}_{1}\right) {}_{i \, {\hat k}}  ~&=~ \left[\begin{array}{cccc}
~1 & ~0 &  ~0  &  ~0 \\
~0 & ~0 &  ~0  &  - \, 1 \\
~\ell_{3 \,1} & ~\ell_{3 \,2} &   \ell_{3 \,3}  &  \ell_{3 \,4} \\
~\ell_{4 \,1} & ~\ell_{4 \,2} &  ~ \ell_{4 \,3}  &   \ell_{4 \,4} \\
\end{array}\right]   ~~, ~~
\left( {\rm L}{}_{2}\right) {}_{i \, {\hat k}} ~=~ \left[\begin{array}{cccc}
\, \, 0 & ~1 &  ~0  &  ~0 \,  \\
\, \, 0 & ~0 &  ~1  &  ~0 \\
-\, \ell_{3 \,2} & \ell_{3 \,1} &   - \ell_{3 \,4}  &   \ell_{3 \,3} \,  \, \\
- \, \ell_{4 \,2} & \ell_{4 \,1} &   - \ell_{4 \,4}  &  ~ \ell_{4 \,3} \\
\end{array}\right]   ~~~,  \cr
\left( {\rm L}{}_{3}\right) {}_{i \, {\hat k}} ~&=~ \left[\begin{array}{cccc}
0 & ~0 &  ~1  &  ~0 \\
0 & -\, 1 &  ~0  &  ~0 \\
 - \ell_{3 \,3} &  \ell_{3 \,4} &   \ell_{3 \,1}  &  - \, \ell_{3 \,2} \\
- \ell_{4 \,3} &  \ell_{4 \,4} &   \ell_{4 \,1}  &  - \,  \ell_{4 \,2} \\
\end{array}\right]   ~, ~
\left( {\rm L}{}_{4}\right) {}_{i \, {\hat k}} ~=~ \left[\begin{array}{cccc}
0 & 0 &  ~0  &  ~1 \\
1 & 0 &  ~0  &  ~0 \\
 - \ell_{3 \,4} &  - \ell_{3 \,3} &   \ell_{3 \,2}  &  ~ \ell_{3 \,1} \\
 - \ell_{4 \,4} & -  \ell_{4 \,3} &   \ell_{4 \,2}  &  ~ \ell_{4 \,1} \, \,  \\
\end{array}\right]  ~~,  \label{MtrXL}
} \ee
We find that a corresponding set of R-matrices to satisfy (\ref{GDNAlge1SPLT1}) is
given by
\be   \eqalign{
\left( {\rm R}{}_{1} \right) {}_{{\hat k} \, i}  ~&=~ \left[\begin{array}{cccc}
~1 & ~0 & \ell_{3 \, 1}  &  \ell_{4 \, 1} \\
~0 & ~0 & \ell_{3 \, 2}  &  \ell_{4 \, 2} \\
~0 & ~0 & \ell_{3 \, 3}  &  \ell_{4 \, 3} \\
~0 & - 1 & \ell_{3 \, 4}  &  \ell_{4 \, 4} \\
\end{array}\right]     ~~~~~~~~ ,  ~~~~
\left( {\rm R}{}_{2} \right) {}_{{\hat k} \, i} ~=~ \left[\begin{array}{cccc}
~0 & ~0 & - \ell_{3 \, 2}  &  -\ell_{4 \, 2} \\
~1 & ~0 & \ell_{3 \, 1}  &  \ell_{4 \, 1} \\
~0 & ~1 & - \ell_{3 \, 4}  &  - \ell_{4 \, 4} \\
~0 & ~0 & \ell_{3 \, 3}  &  \ell_{4 \, 3} \\
\end{array}\right]   ~~, ~~  \cr
\left( {\rm R}{}_{3} \right) {}_{{\hat k} \, i} ~&=~ \left[\begin{array}{cccc}
~0 & ~0 & - \ell_{3 \, 3}  & - \ell_{4 \, 3} \\
~0 & - 1 & \ell_{3 \, 4}  &  \ell_{4 \, 4} \\
~1 & ~0 & \ell_{3 \, 1}  &  \ell_{4 \, 1} \\
~0 & ~0 & - \ell_{3 \, 2}  &  - \ell_{4 \, 2} \\
\end{array}\right]  ~~~~, ~~~~ 
\left( {\rm R}{}_{4} \right) {}_{{\hat k} \, i}  =~ \left[\begin{array}{cccc}
~0 & ~1 & - \ell_{3 \, 4}  & - \ell_{4 \, 4} \\
~0 & ~0 & - \ell_{3 \, 3}  & - \ell_{4 \, 3} \\
~0 & ~0 & \ell_{3 \, 2}  &  \ell_{4 \, 2} \\
~1 & ~0 & \ell_{3 \, 1}  &  \ell_{4 \, 1} \\
\end{array}\right]~~~,  \label{MtrXL2}
} \ee
However, we can 
use these augmented L-matrices and R-matrices to carry out the calculations
indicated in (\ref{GDNAlge1SPLT2}).  For these calculations 
we find
\be \eqalign{
 (\,{\rm R}_1\,)_\hi{}^j\>(\, {\rm L}_2\,)_j{}^\hk + (\,{\rm R}_2\,)_\hi{}^j\>(\,{\rm 
 L}_1\,)_j{}^\hk
   ~= {~~~~~~~~~~~~~~~~~~~~~~~~~~~~~~~~~~~~}&
   \cr
{~~~~}  
\left[\begin{array}{cccc}
- 2 {\cal P}^{[1|2]}_{1\, 1} & 1 + {\cal P}^{[1|2]}_{1\, 2} & - {\cal P}^{[1|2]}_{1\, 3} & {\cal P}^{[1|2]}_{1\, 4} \\
1 + {\cal P}^{[1|2]}_{1\, 2} &  2 {\cal P}^{[1|2]}_{1\, 1} & {\cal P}^{[1|2]}_{1\, 4} &  {\cal P}^{[1|2]}_{1\, 3} \\
-   {\cal P}^{[1|2]}_{1\, 3}  &  {\cal P}^{[1|2]}_{1\, 4} &- 2 {\cal P}^{[1|2]}_{3\, 3} & - 1 +  {\cal P}^{[1|2]}_{3\, 4} \\
{\cal P}^{[1|2]}_{1\, 4} &  {\cal P}^{[1|2]}_{1\, 3} & - 1 +  {\cal P}^{[1|2]}_{3\, 4}  & 2 {\cal P}^{[1|2]}_{3\, 3} \\
\end{array}\right]
  ~~,
}  \label{GarDNAlg1XX12}
\ee
where
\be   \eqalign{
{\cal P}^{[1|2]}_{1 \,1} &=~  \ell _{31} \ell _{32} + \ell _{41} \ell _{42} ~~,~~  
{\cal P}^{[1|2]}_{1\, 2} ~=~ \ell _{31}^2-\ell _{32}^2+\ell _{41}^2-\ell _{42}^2  ~~,  \cr
{\cal P}^{[1|2]}_{1 \,3} &=~  \ell _{32} \ell _{33}+\ell_{31} \ell _{34}+\ell _{42} \ell _{43}+\ell _{41} \ell _{44}
  ~~,  \cr
{\cal P}^{[1|2]}_{1 \,4} &=~  \ell _{31} \ell _{33}-\ell _{32} \ell _{34}+\ell _{41} \ell _{43}-\ell _{42} \ell _{44}
~~,  \cr
{\cal P}^{[1|2]}_{3 \,3} &=~  \ell _{33} \ell _{34}+ \ell _{43} \ell _{44}  ~~,  ~~
{\cal P}^{[1|2]}_{3 \,4} ~=~ \ell _{33}^2-\ell _{34}^2+\ell _{43}^2-\ell _{44}^2 ~~.
} \label{W1}
\ee

\be \eqalign{
 (\,{\rm R}_1\,)_\hi{}^j\>(\, {\rm L}_3\,)_j{}^\hk + (\,{\rm R}_3\,)_\hi{}^j\>(\,{\rm 
 L}_1\,)_j{}^\hk
    ~= {~~~~~~~~~~~~~~~~~~~~~~~~~~~~~~~~~~~~}&
   \cr
{~~~~}  
\left[\begin{array}{cccc}
- 2 {\cal P}^{[1|3]}_{1\, 1} & - {\cal P}^{[1|3]}_{1\, 2} & 1 + {\cal P}^{[1|3]}_{1\, 3} & - {\cal P}^{[1|3]}_{1\, 4} \\
- {\cal P}^{[1|3]}_{1\, 2} & 2 {\cal P}^{[1|3]}_{2\, 2} & {\cal P}^{[1|3]}_{1\, 4} & 1 - {\cal P}^{[1|3]}_{2\, 4} \\
1 + {\cal P}^{[1|3]}_{1\, 3} & {\cal P}^{[1|3]}_{1\, 4} & 2 {\cal P}^{[1|3]}_{1\, 1} & - {\cal P}^{[1|3]}_{1\, 2} \\
- {\cal P}^{[1|3]}_{1\, 4} & 1 - {\cal P}^{[1|3]}_{2\, 4} & - {\cal P}^{[1|3]}_{1\, 2} & - 2 {\cal P}^{[1|3]}_{2\, 2}
\end{array}\right] ~~~,
}  \label{GarDNAlg1XX13}
\ee
where
\be \eqalign{
{\cal P}^{[1|3]}_{1\, 1} &=  \ell _{31} \ell _{33} + \ell _{41} \ell _{43}
~~,~~ 
{\cal P}^{[1|3]}_{1\, 3} = \ell _{31}^2-\ell _{33}^2+\ell _{41}^2-\ell _{43}^2  ~~,
\cr
 {\cal P}^{[1|3]}_{1\, 2} &= \ell _{32} \ell _{33}-\ell _{31} \ell _{34}+\ell _{42} \ell _{43}-\ell_{41} \ell _{44} ~~,
\cr
{\cal P}^{[1|3]}_{1\, 4} &=  \ell _{31} \ell _{32}+\ell _{33} \ell _{34}+\ell _{41} \ell_{42}+\ell _{43} \ell _{44} ~~,
\cr
{\cal P}^{[1|3]}_{2\, 2} &=  \ell _{32} \ell _{34}+ \ell _{42} \ell _{44}
~~,~~ 
{\cal P}^{[1|3]}_{2\, 4} = \ell_{32}^2-\ell _{34}^2+\ell _{42}^2-\ell _{44}^2 ~~,
\cr
}  \label{W2}
\ee

\be \eqalign{
 (\,{\rm R}_1\,)_\hi{}^j\>(\, {\rm L}_4\,)_j{}^\hk + (\,{\rm R}_4\,)_\hi{}^j\>(\,{\rm 
 L}_1\,)_j{}^\hk
    ~= {~~~~~~~~~~~~~~~~~~~~~~~~~~~~~~~~~~~~}&
   \cr
{~~~~}  
\left[\begin{array}{cccc}
- 2 {\cal P}^{[1|4]}_{1\, 1} &
- {\cal P}^{[1|4]}_{1\, 2}  &
 {\cal P}^{[1|4]}_{1\, 3} & 
  {\cal P}^{[1|4]}_{1\, 4} \\
- {\cal P}^{[1|4]}_{1\, 2}  &
- 2 {\cal P}^{[1|4]}_{2\, 2}  &
{\cal P}^{[1|4]}_{2\, 3}  &
{\cal P}^{[1|4]}_{1\, 3} \\ 
{\cal P}^{[1|4]}_{1\, 3} & 
{\cal P}^{[1|4]}_{2\, 3}  &
2 {\cal P}^{[1|4]}_{2\, 2}  &
{\cal P}^{[1|4]}_{1\, 2}  \\
{\cal P}^{[1|4]}_{1\, 4} &
{\cal P}^{[1|4]}_{1\, 3} & 
{\cal P}^{[1|4]}_{1\, 2}  &
2 {\cal P}^{[1|4]}_{1\, 1} 
\end{array}\right] ~~~,
}  \label{GarDNAlg1XX14}
\ee
where
\be  \eqalign{
{\cal P}^{[1|4]}_{1\, 1} &=  \ell _{31} \ell _{34}+ \ell _{41} \ell _{44}
~,~
{\cal P}^{[1|4]}_{1\, 4} = 
\ell _{31}^2-\ell_{34}^2+\ell _{41}^2-\ell _{44}^2  ~~, \cr
{\cal P}^{[1|4]}_{1\, 2} &= 
 \ell _{31} \ell _{33}+\ell _{32} \ell _{34}+\ell _{41} \ell _{43}+\ell_{42} \ell _{44} ~~,
\cr
{\cal P}^{[1|4]}_{1\, 3} &= 
 \ell _{31} \ell _{32}-\ell _{33} \ell _{34}+\ell _{41} \ell _{42}-\ell _{43} \ell _{44}
 ~~, \cr
 {\cal P}^{[1|4]}_{2\, 2} &= 
 \ell _{32} \ell _{33}+ \ell _{42} \ell _{43}
 ~,~
 {\cal P}^{[1|4]}_{2\, 3} = 
  \ell _{32}^2-\ell _{33}^2+\ell _{42}^2-\ell _{43}^2 ~~,
}  \label{W3}
\ee

\be \eqalign{
 (\,{\rm R}_2\,)_\hi{}^j\>(\, {\rm L}_3\,)_j{}^\hk + (\,{\rm R}_3\,)_\hi{}^j\>(\,{\rm 
 L}_2\,)_j{}^\hk
     ~= {~~~~~~~~~~~~~~~~~~~~~~~~~~~~~~~~~~~~}&
   \cr
{~~~~}  
\left[\begin{array}{cccc}
2 {\cal P}^{[2|3]}_{1\, 1} &
- {\cal P}^{[2|3]}_{1\, 2} &
- {\cal P}^{[2|3]}_{1\, 3} &
{\cal P}^{[2|3]}_{1\, 4} \\
- {\cal P}^{[2|3]}_{1\, 2} &
2  {\cal P}^{[2|3]}_{2\, 2} &
{\cal P}^{[2|3]}_{2\, 3} &
- {\cal P}^{[2|3]}_{1\, 3} \\
- {\cal P}^{[2|3]}_{1\, 3} &
{\cal P}^{[2|3]}_{2\, 3} &
- 2  {\cal P}^{[2|3]}_{2\, 2} &
{\cal P}^{[2|3]}_{1\, 2} \\
{\cal P}^{[2|3]}_{1\, 4} &
- {\cal P}^{[2|3]}_{1\, 3} &
 {\cal P}^{[2|3]}_{1\, 2} &
 - 2 {\cal P}^{[2|3]}_{1\, 1}
\end{array}\right] ~~~,
}  \label{GarDNAlg1XX23}
\ee
where
\be \eqalign{
{\cal P}^{[2|3]}_{1\, 1} &= \ell _{32} \ell _{33}+ \ell _{42} \ell _{43} 
~,~
{\cal P}^{[2|3]}_{1\, 4} = \ell_{32}^2-\ell _{33}^2+\ell _{42}^2-\ell _{43}^2 ~,~
\cr
{\cal P}^{[2|3]}_{1\, 2} &=  \ell _{31} \ell _{33}+\ell _{32} \ell _{34}+\ell _{41} \ell _{43}+\ell_{42} \ell _{44}
~,~
\cr
{\cal P}^{[2|3]}_{1\, 3} &=   \ell _{31} \ell _{32}-\ell _{33} \ell _{34}+\ell _{41} \ell _{42}-\ell _{43} \ell _{44}
~,~
\cr
{\cal P}^{[2|3]}_{2\, 2} &= \ell _{31} \ell _{34}+ \ell _{41} \ell _{44}
~,~
{\cal P}^{[2|3]}_{2\, 3} =  \ell _{31}^2-\ell _{34}^2+\ell _{41}^2-\ell _{44}^2 ~.
}  \label{W4}
\ee

\be \eqalign{
(\,{\rm R}_2\,)_\hi{}^j\>(\, {\rm L}_4\,)_j{}^\hk + (\,{\rm R}_4\,)_\hi{}^j\>(\,{\rm 
L}_2\,)_j{}^\hk
    ~= {~~~~~~~~~~~~~~~~~~~~~~~~~~~~~~~~~~~~}&
   \cr
\left[\begin{array}{cccc}
2 {\cal P}^{[2|4]}_{1\, 1} &
{\cal P}^{[2|4]}_{1\, 2} &
1 - {\cal P}^{[2|4]}_{1\, 3} &
- {\cal P}^{[2|4]}_{1\, 4} \\
{\cal P}^{[2|4]}_{1\, 2} &
- 2 {\cal P}^{[2|4]}_{2\, 2} &
 {\cal P}^{[2|4]}_{1\, 4} &
 1 + {\cal P}^{[2|4]}_{2\, 4} \\
1 - {\cal P}^{[2|4]}_{1\, 3} &
 {\cal P}^{[2|4]}_{1\, 4} &
- 2 {\cal P}^{[2|4]}_{1\, 1} & 
{\cal P}^{[2|4]}_{1\, 2} \\
- {\cal P}^{[2|4]}_{1\, 4} &
 1 + {\cal P}^{[2|4]}_{2\, 4} &
 {\cal P}^{[2|4]}_{1\, 2} &
 2 {\cal P}^{[2|4]}_{2\, 2}
\end{array}\right] ~~~,
}  \label{GarDNAlg1XX24}
\ee
where
\be  \eqalign{
{\cal P}^{[2|4]}_{1\, 1} &= \ell _{32} \ell _{34}+ \ell _{42} \ell _{44} ~,~ 
{\cal P}^{[2|4]}_{1\, 3} = \ell _{32}^2-\ell _{34}^2+\ell _{42}^2-\ell _{44}^2 ~~, \cr
{\cal P}^{[2|4]}_{1\, 2} &=  \ell _{32} \ell _{33}-\ell _{31} \ell _{34}+\ell _{42} \ell _{43}-\ell_{41} \ell _{44} ~,~ \cr
{\cal P}^{[2|4]}_{1\, 4} &=
 \ell _{31} \ell _{32}+\ell _{33} \ell _{34}+\ell _{41} \ell _{42}+\ell _{43} \ell _{44} ~,~ \cr
 {\cal P}^{[2|4]}_{2\, 2} &=  \ell _{31} \ell _{33}+ \ell _{41} \ell _{43} ~,~
  {\cal P}^{[2|4]}_{2\, 4} =
 \ell_{31}^2-\ell _{33}^2+\ell _{41}^2-\ell _{43}^2 ~~,
}  \label{W5}
\ee

\be \eqalign{
 (\,{\rm R}_3\,)_\hi{}^j\>(\, {\rm L}_4\,)_j{}^\hk + (\,{\rm R}_4\,)_\hi{}^j\>(\,{\rm 
 L}_3\,)_j{}^\hk
     ~= {~~~~~~~~~~~~~~~~~~~~~~~~~~~~~~~~~~~~}&
   \cr
{~~~~}  
\left[\begin{array}{cccc}
2 {\cal P}^{[3|4]}_{1 \,1}  &
- 1 +  {\cal P}^{[3|4]}_{1 \,2}  &
-  {\cal P}^{[3|4]}_{1 \,3}  &
- {\cal P}^{[3|4]}_{1 \,4}  \\
 - 1 +  {\cal P}^{[3|4]}_{1 \,2}  &
 - 2 {\cal P}^{[3|4]}_{1 \,1}  &
-  {\cal P}^{[3|4]}_{1 \,4}  &
 {\cal P}^{[3|4]}_{1 \,3}  \\
 -  {\cal P}^{[3|4]}_{1 \,3}  &
- {\cal P}^{[3|4]}_{1 \,4}  &
2 {\cal P}^{[3|4]}_{3 \,3}  &
 1 +  {\cal P}^{[3|4]}_{3 \,4}  \\
 -  {\cal P}^{[3|4]}_{1 \,4}  &
 {\cal P}^{[3|4]}_{1 \,3}  &
1 +  {\cal P}^{[3|4]}_{3 \,4}  &
- 2 {\cal P}^{[3|4]}_{3 \,3}
\end{array}\right] ~~~,
}  \label{GarDNAlg1XX34}
\ee
where
\be
\eqalign{
 {\cal P}^{[3|4]}_{1 \,1} &=  \ell _{33} \ell _{34}+ \ell _{43} \ell _{44}
 ~,~
 {\cal P}^{[3|4]}_{1 \,2} = \ell _{33}^2-\ell _{34}^2+\ell _{43}^2-\ell _{44}^2
 ~,~ \cr
  {\cal P}^{[3|4]}_{1 \,3} &=
 \ell _{32} \ell _{33}+\ell _{31} \ell _{34}+\ell _{42} \ell _{43}+\ell _{41} \ell _{44}
 ~~,  \cr
  {\cal P}^{[3|4]}_{1 \,4} &=
\ell _{31} \ell _{33}-\ell _{32} \ell_{34}+\ell _{41} \ell _{43}-\ell _{42} \ell _{44}
~~, \cr
 {\cal P}^{[3|4]}_{3 \,3} &=  \ell _{31} \ell _{32}+ \ell _{41} \ell _{42} ~,~
  {\cal P}^{[3|4]}_{3 \,4} = \ell _{31}^2-\ell _{32}^2+\ell _{41}^2-\ell _{42}^2 
  ~~.
}   \label{W6}
\ee

Imposing the conditions that the matrices in (\ref{GarDNAlg1XX12}),
(\ref{GarDNAlg1XX13}), (\ref{GarDNAlg1XX14}), (\ref{GarDNAlg1XX23}),
(\ref{GarDNAlg1XX24}), and (\ref{GarDNAlg1XX34}) should vanish yields
solutions to these equations
\be
\ell_{3 2} = \pm 1 ~~~,~~~ \ell_{4 3} = \pm 1 ~~~,
\label{ellCondc}
\ee
and all other $\ell$-parameters vanish.
Up to the field redefinitions $F$ $\to$ $- F$ and $G$ $\to$ $- G$,
we have recovered the off-shell chiral adinkra network L-matrices
and R-matrices by starting from the on-shell chiral adinkra network L-matrices
and R-matrices.  The solution in (\ref{ellCondc}) also can be shown to satisfy
 the conditions in (\ref{GDNAlge1SPLT3}) and (\ref{GDNAlge1SPLT4}).

\subsection{On-Shell Vector Valise Matrix Deformation} 
$~~~$ 
Now we repeat the analysis of the previous subsection but switching our
attention to the vector supermultiplet adinkra valise matrices.  We introduce 
a set of deformation to the on-shell L-matrices and R-matrices
shown in (\ref{V1D0E2}) and (\ref{V1D0HOz}) by introducing the deforming parameters
$\ell_{4 \,1}$,  $\ell_{4 \,2}$, $\ell_{4 \,3}$, and $\ell_{4 \,4}$ to augment
the on-shell matrices according to   
$$
\left( {\rm L}{}_{1}\right) {}_{i \, {\hat k}}   ~=~
\left[\begin{array}{cccc}
~0 & ~1 &  ~ 0  &  ~ 0 \\
~0 & ~0 &  ~0  &  -\,1 \\
~1 & ~0 &  ~ 0  &  ~0 \\
 \ell_{4 \,1} & ~ \ell_{4 \,2} & \ell_{4 \,3} & \ell_{4 \,4}
\end{array}\right] ~~~,~~~
\left( {\rm L}{}_{2}\right) {}_{i \, {\hat k}}   ~=~
\left[\begin{array}{cccc}
1 &  \, 0 &  0  &  ~ 0 \\
0 &  \, 0 &  1  &  ~ 0 \\
0 & - \, 1 &  0  &   ~ 0 \\
\ell_{4 \,2} & - \ell_{4 \,1} &  - \ell_{4 \,4}  &  \ell_{4 \,3}
\end{array}\right]  ~~~,
$$
\be 
\left( {\rm L}{}_{3}\right) {}_{i \, {\hat k}}   ~=~
\left[\begin{array}{cccc}
0 & \,0 &  \, 0  &  ~ 1 \\
0 & \,1 &  \,0  &   ~0 \\
0 & \,0 &  \, 1  &  ~0 \\
- \ell_{4 \,3} & - \ell_{4 \,4} &  \ell_{4 \,1}  &   \ell_{4 \,2}
\end{array}\right] ~~~~,~~~
\left( {\rm L}{}_{4}\right) {}_{i \, {\hat k}}   ~=~
\left[\begin{array}{cccc}
~0 & ~0 &  ~1  &  ~ 0 \\
-\,1 & ~ 0 &  ~0  &  ~ 0 \\
 ~0 & ~0 &  ~0  &   - \, 1 \\
\ell_{4 \,4} & - \ell_{4 \,3} &   \ell_{4 \,2}  &   - \ell_{4 \,1}
\end{array}\right]  ~~~,
\label{V1D0Y1}
\ee
$$
\left( {\rm R}{}_{1}\right) {}_{{\hat k} \, i}   ~=~
\left[\begin{array}{cccc}
~0 & ~0 &  ~ 1  &  ~ \ell_{4\, 1} \\
~1 & ~0 &  ~0  &  ~ \ell_{4\, 2} \\
~0 & ~0 &  ~0   &  ~ \ell_{4\, 3}  \\
~0 & -\, 1 &  ~0  &  ~ \ell_{4\, 4}
\end{array}\right] ~~~,~~~
\left( {\rm R}{}_{2}\right) {}_{{\hat k} \, i}   ~=~
\left[\begin{array}{cccc}
~1 & ~ 0 &  ~0  &  ~ \ell_{4\, 2} \\
~0 & ~ 0 &  -\,1  &  ~ - \ell_{4\, 1} \\
 ~0 & ~ 1 &  ~0  &   ~ - \ell_{4\, 4} \\
 ~0 & ~0 &  ~0  &   ~  \ell_{4\, 3} \\
\end{array}\right]  ~~,~ ~~~~
$$
\be 
\left( {\rm R}{}_{3}\right) {}_{{\hat k} \, i}   ~=~
\left[\begin{array}{cccc}
~0 & ~0 &  ~0  &   - \ell_{4\, 3}\\
~0 & ~1 &  ~0  &   - \ell_{4\, 4} \\
~0 & ~0 &  ~ 1  &  ~\ell_{4\, 1} \\
~1 & ~0&  ~0  &  ~ \ell_{4\, 2}
\end{array}\right] ~~~~~~,~~~
\left( {\rm R}{}_{4}\right) {}_{{\hat k} \, i}  ~=~
\left[\begin{array}{cccc}
~0 & - \, 1 &  ~ 0  &  ~ \ell_{4\, 4} \\
~0 & `0 &  ~ 0  &  -  \ell_{4\, 3} \\
~1 & ~ 0 &  ~ 0  &  ~ \ell_{4\, 2} \\
 ~0 & ~ 0 &  - \, 1  &  - \, \ell_{4\, 1} \\
\end{array}\right]  ~~~,
\label{V1D0Y2}
\ee
Once more direct calculations show these satisfy (\ref{GDNAlge1SPLT1}).  
However, we can also carry out similar calculations where the R-matrices
appearing as the terms farthest to the left in the matrix multiplications.  
For these calculations 
we find
\be \eqalign{
 (\,{\rm R}_1\,)_\hi{}^j\>(\, {\rm L}_2\,)_j{}^\hk + (\,{\rm R}_2\,)_\hi{}^j\>(\,{\rm 
 L}_1\,)_j{}^\hk
  ~= {~~~~~~~~~~~~~~~~~~~~~~~~~~~~~~~~~~~~~~~~~~~~~~~~~~~~}&
   \cr
{~~~~}  \left[\begin{array}{cccc}
 2 \ell _{41} \ell _{42} & -\ell _{41}^2+\ell _{42}^2 & \ell _{42} \ell _{43}-\ell _{41} \ell _{44} & \ell _{41} \ell
_{43}+\ell _{42} \ell _{44} \\
 -\ell _{41}^2+\ell _{42}^2 & -2 \ell _{41} \ell _{42} & -\ell _{41} \ell _{43}-\ell _{42} \ell _{44} & \ell _{42} \ell
_{43}-\ell _{41} \ell _{44} \\
 \ell _{42} \ell _{43}-\ell _{41} \ell _{44} & -\ell _{41} \ell _{43}-\ell _{42} \ell _{44} & -2 \ell _{43} \ell _{44}
& -1+\ell _{43}^2-\ell _{44}^2 \\
 \ell _{41} \ell _{43}+\ell _{42} \ell _{44} & \ell _{42} \ell _{43}-\ell _{41} \ell _{44} & -1+\ell _{43}^2-\ell _{44}^2
& 2 \ell _{43} \ell _{44}
\end{array}\right] 
  ~~~,
}  \label{GarDNAlg1YY12}
\ee
\be \eqalign{
 (\,{\rm R}_1\,)_\hi{}^j\>(\, {\rm L}_3\,)_j{}^\hk + (\,{\rm R}_3\,)_\hi{}^j\>(\,{\rm 
 L}_1\,)_j{}^\hk
 ~= {~~~~~~~~~~~~~~~~~~~~~~~~~~~~~~~~~~~~~~~~~~~~~~~~~~~~}&
 \cr
{~~~~}  \left[\begin{array}{cccc}   
 -2 \ell _{41} \ell _{43} & -\ell _{42} \ell _{43}-\ell _{41} \ell _{44} & 1+\ell _{41}^2-\ell _{43}^2 & \ell _{41} \ell
_{42}-\ell _{43} \ell _{44} \\
 -\ell _{42} \ell _{43}-\ell _{41} \ell _{44} & -2 \ell _{42} \ell _{44} & \ell _{41} \ell _{42}-\ell _{43} \ell _{44}
& \ell _{42}^2-\ell _{44}^2 \\
 1+\ell _{41}^2-\ell _{43}^2 & \ell _{41} \ell _{42}-\ell _{43} \ell _{44} & 2 \ell _{41} \ell _{43} & \ell _{42} \ell
_{43}+\ell _{41} \ell _{44} \\
 \ell _{41} \ell _{42}-\ell _{43} \ell _{44} & \ell _{42}^2-\ell _{44}^2 & \ell _{42} \ell _{43}+\ell _{41} \ell _{44}
& 2 \ell _{42} \ell _{44}
\end{array}\right]  ~~~,
}  \label{GarDNAlg1YY13}
\ee
\be \eqalign{
 (\,{\rm R}_1\,)_\hi{}^j\>(\, {\rm L}_4\,)_j{}^\hk + (\,{\rm R}_4\,)_\hi{}^j\>(\,{\rm 
 L}_1\,)_j{}^\hk
  ~= {~~~~~~~~~~~~~~~~~~~~~~~~~~~~~~~~~~~~~~~~~~~~~~~~~~~~}&
 \cr
{~~~~} 
\left[\begin{array}{cccc}  
 2 \ell _{41} \ell _{44} & -\ell _{41} \ell _{43}+\ell _{42} \ell _{44} & \ell _{41} \ell _{42}+\ell _{43} \ell _{44}
& -\ell _{41}^2+\ell _{44}^2 \\
 -\ell _{41} \ell _{43}+\ell _{42} \ell _{44} & -2 \ell _{42} \ell _{43} & 1+\ell _{42}^2-\ell _{43}^2 & -\ell _{41} \ell
_{42}-\ell _{43} \ell _{44} \\
 \ell _{41} \ell _{42}+\ell _{43} \ell _{44} & 1+\ell _{42}^2-\ell _{43}^2 & 2 \ell _{42} \ell _{43} & -\ell _{41} \ell
_{43}+\ell _{42} \ell _{44} \\
 -\ell _{41}^2+\ell _{44}^2 & -\ell _{41} \ell _{42}-\ell _{43} \ell _{44} & -\ell _{41} \ell _{43}+\ell _{42} \ell
_{44} & -2 \ell _{41} \ell _{44}
\end{array}\right]  \, , {~~~~~}
}  \label{GarDNAlg1YY14}
\ee
\be \eqalign{
 (\,{\rm R}_2\,)_\hi{}^j\>(\, {\rm L}_3\,)_j{}^\hk + (\,{\rm R}_3\,)_\hi{}^j\>(\,{\rm 
 L}_2\,)_j{}^\hk
  ~= {~~~~~~~~~~~~~~~~~~~~~~~~~~~~~~~~~~~~~~~~~~~~~~~~~~~~}&
 \cr
{~~~~} 
\left[\begin{array}{cccc} 
 -2 \ell _{42} \ell _{43} & \ell _{41} \ell _{43}-\ell _{42} \ell _{44} & \ell _{41} \ell _{42}+\ell _{43} \ell _{44}
& 1+\ell _{42}^2-\ell _{43}^2 \\
 \ell _{41} \ell _{43}-\ell _{42} \ell _{44} & 2 \ell _{41} \ell _{44} & -\ell _{41}^2+\ell _{44}^2 & -\ell _{41} \ell
_{42}-\ell _{43} \ell _{44} \\
 \ell _{41} \ell _{42}+\ell _{43} \ell _{44} & -\ell _{41}^2+\ell _{44}^2 & -2 \ell _{41} \ell _{44} & \ell _{41} \ell
_{43}-\ell _{42} \ell _{44} \\
 1+\ell _{42}^2-\ell _{43}^2 & -\ell _{41} \ell _{42}-\ell _{43} \ell _{44} & \ell _{41} \ell _{43}-\ell _{42} \ell
_{44} & 2 \ell _{42} \ell _{43}
\end{array}\right]   \, , {~~~~~}
}  \label{GarDNAlg1YY23}
\ee
\be \eqalign{
(\,{\rm R}_2\,)_\hi{}^j\>(\, {\rm L}_4\,)_j{}^\hk + (\,{\rm R}_4\,)_\hi{}^j\>(\,{\rm 
L}_2\,)_j{}^\hk
  ~= {~~~~~~~~~~~~~~~~~~~~~~~~~~~~~~~~~~~~~~~~~~~~~~~~~~~~}&
 \cr
{~~~~} 
\left[\begin{array}{cccc} 
 2 \ell _{42} \ell _{44} & -\ell _{42} \ell _{43}-\ell _{41} \ell _{44} & \ell _{42}^2-\ell _{44}^2 & -\ell _{41} \ell
_{42}+\ell _{43} \ell _{44} \\
 -\ell _{42} \ell _{43}-\ell _{41} \ell _{44} & 2 \ell _{41} \ell _{43} & -\ell _{41} \ell _{42}+\ell _{43} \ell _{44}
& 1+\ell _{41}^2-\ell _{43}^2 \\
 \ell _{42}^2-\ell _{44}^2 & -\ell _{41} \ell _{42}+\ell _{43} \ell _{44} & -2 \ell _{42} \ell _{44} & \ell _{42} \ell
_{43}+\ell _{41} \ell _{44} \\
 -\ell _{41} \ell _{42}+\ell _{43} \ell _{44} & 1+\ell _{41}^2-\ell _{43}^2 & \ell _{42} \ell _{43}+\ell _{41} \ell
_{44} & -2 \ell _{41} \ell _{43}
\end{array}\right]   \, , {~~~~~}
}  \label{GarDNAlg1YY24}
\ee
\be \eqalign{
 (\,{\rm R}_3\,)_\hi{}^j\>(\, {\rm L}_4\,)_j{}^\hk + (\,{\rm R}_4\,)_\hi{}^j\>(\,{\rm 
 L}_3\,)_j{}^\hk
   ~= {~~~~~~~~~~~~~~~~~~~~~~~~~~~~~~~~~~~~~~~~~~~~~~~~~~~~}&
 \cr
{~~~~} 
\left[\begin{array}{cccc} 
 -2 \ell _{43} \ell _{44} & -1+\ell _{43}^2-\ell _{44}^2 & -\ell _{42} \ell _{43}+\ell _{41} \ell _{44} & \ell _{41} \ell
_{43}+\ell _{42} \ell _{44} \\
 -1+\ell _{43}^2-\ell _{44}^2 & 2 \ell _{43} \ell _{44} & -\ell _{41} \ell _{43}-\ell _{42} \ell _{44} & -\ell _{42} \ell
_{43}+\ell _{41} \ell _{44} \\
 -\ell _{42} \ell _{43}+\ell _{41} \ell _{44} & -\ell _{41} \ell _{43}-\ell _{42} \ell _{44} & 2 \ell _{41} \ell _{42}
& -\ell _{41}^2+\ell _{42}^2 \\
 \ell _{41} \ell _{43}+\ell _{42} \ell _{44} & -\ell _{42} \ell _{43}+\ell _{41} \ell _{44} & -\ell _{41}^2+\ell _{42}^2
& -2 \ell _{41} \ell _{42}
\end{array}\right]   \, . {~~~~~}
}  \label{GarDNAlg1YY34}
\ee
If we impose the condition in (\ref{GDNAlge1SPLT2}) we are easily 
led to the soluitions 
\be \eqalign{
\ell_{4\, 1} ~=~  \ell_{4\, 2} ~=~  \ell_{4\, 4} ~=~ 0 ~~~,~~~  \ell_{4\, 3} ~=~ \pm 1 ~~~.
}  \label{ellCondv}
\ee
Up to a sign (which corresponds to the redefinition 
d $\to$ $-$ d) we recover the off-shell L-matrices and R-matrices of (\ref{V1D0E}) 
and (\ref{V1D0HO}) for the adinkra network version 
of vector supermultiplet.  The solution in (\ref{ellCondv}) also can be shown to satisfy
 the conditions in (\ref{GDNAlge1SPLT3}) and (\ref{GDNAlge1SPLT4}).

So once again we see the method of deforming the on-shell matrices by augmentation
involving the $\ell$-parameters followed by the imposition of the off-diagonal
part of the  Garden Algebra conditions leads from the on-shell to the off-shell versions 
of the matrices.
To summarize the results of this chapter, we have shown that one can:
\newline \indent
(a.) start with on-shell L-matrices and R-matrices (for the chiral adinkra 
\newline \indent $~~~~~$
network (\ref{chiD0N}) and (\ref{chiD0O}) or for the vector adinkra
network (\ref{V1D0E2}) and (\ref{V1D0HOz})),
\newline \indent
(b.) use $\ell$-parameters to augment the on-shell L-matrices and R-matrices
\newline \indent $~~~~~$ (for the chiral adinkra network (\ref{MtrXL}) and
(\ref{MtrXL2}) or for the vector adinkra \newline \indent $~~~~~$ 
network (\ref{V1D0Y1}) and (\ref{V1D0Y2})),
\newline \indent
(c.) impose the Garden Algebra conditions in (\ref{GDNAlge1SPLT1}) and
(\ref{GDNAlge1SPLT2}), and 
\newline \indent
(d.) thereby, up to a set of field redefinitions, derive the off-shell versions
\newline \indent $~~~~~$ of the respective L-matrices and R-matrices.
(for the chiral adinkra \newline \indent $~~~~~$
network (\ref{chiD0F}) and (\ref{chiD0K})
or for the vector adinkra (\ref{V1D0E}) and (\ref{V1D0HO})). 
\vskip.03in \noindent
The $\ell$-augmented L-matrices and R-matrices interpolate between the
on-shell solutions (where all $\ell$ parameters vanish) and the off-shell ones
(where the $\ell$ parameters take on the values shown in (\ref{ellCondc})
or (\ref{ellCondv}) in the respective cases).  For general values of the $\ell$-parameters,
the augmented matrices do not satisfy the Garden Algebra.

\section{The General Cryptographic Problem Analogy to the 
Adinkra Network Auxiliary Field Problem} 
$~~~$
In this chapter, we want to discuss the general matrix problem that
adinkra networks provide as the translation of the off-shell SUSY
auxiliary field problem.

Consider a set of matrices of the forms
\be
 (\,{\rm L}_\rI\,)_{i \, {\hat k}} ~=~
\left[\begin{array}{ccccccc} 
a^{\rm I}_{1 \, 1}  & a^{\rm I}_{1 \, 2} & \cdots & a^{\rm I}_{1 \, r_1} 
& \ell^{\rm I}_{1 \, r_1 + 1} & \cdots & \ell^{\rm I}_{1 \, 4 p} \\
a^{\rm I}_{2 \, 1}  & a^{\rm I}_{2 \, 2} & \cdots & a^{\rm I}_{2 \, r_2} 
& \ell^{\rm I}_{1 \, r_2 + 1} & \cdots & \ell^{\rm I}_{2 \, 4 p} \\
\vdots & {~} & {~} & {~} & {~} & {~} & \vdots \\
\vdots & {~} & {~} & {~} & {~} & {~} & \vdots \\
\vdots & {~} & {~} & {~} & {~} & {~} & \vdots \\
\vdots & {~} & {~} & {~} & {~} & {~} & \vdots \\
\vdots & {~} & {~} & {~} & {~} & {~} & \vdots \\
a^{\rm I}_{4 p \, 1} & {~} & {~} & \cdots & {~} & {~} & \ell^{\rm I}_{4 p \, 4 p}
\end{array}\right]    ~~,
\ee
and
\be
 (\,{\rm R}_\rI\,)_{{\hat k} \, i} ~=~
\left[\begin{array}{ccccccc} 
b^{\rm I}_{1 \, 1}  & b^{\rm I}_{1 \, 2} & {~} & \cdots & {~} & {~} & b{}^{\rm I}_{1 \, 4 p} \\
b^{\rm I}_{2 \, 1}  & b^{\rm I}_{2 \, 2} & {~} & \cdots & {~} & {~} & b{}^{\rm I}_{2 \, 4 p} \\
\vdots & {~} & {~} & {~} & {~} & {~} & \vdots   \\
b^{\rm I}_{s_1 \, 1}  & b^{\rm I}_{s_1 \, 2} & {~} & \cdots & {~} & {~} & b{}^{\rm I}_{s_1 \, 4 p} \\
{\Hat \ell}{}^{\rm I}_{s_1 +1  \, 1} & {\Hat \ell}{}^{\rm I}_{s_1 +1  \, 2} & {~} & \cdots & {~} & 
{~} & {\Hat \ell}{}^{\rm I}_{s_1 +1 \, 4 p} \\
\vdots & {~~} & {~~} & {~~} & {~~} & {~~\,~} & \vdots  \\
\vdots & {~} & {~} & {~} & {~} & {~} & \vdots \\
{\Hat \ell}{}^{\rm I}_{4 p \, 1} & {~} & {~} & \cdots & {~} & {~} & {\Hat \ell}{}^{\rm I}_{4 p \, 4 p}
\end{array}\right]   ~~.
\ee
with I = 1 $\dots$ $N$.  In writing these expressions the integer $p$ is assumed to be
some fixed counting number.  The integers $r_1$ to $r_{4p}$ are allowed to range from
$0$ to $4p - 1$ and similarly the integers $s_1$ to $s_{4p}$ are allowed to range from
$0$ to $4p - 1$.  We also assume that the numerical values of all the entries in the
matrices are such that they satisfy the constraints in satisfy the conditions in  (\ref{GDNAlge1SPLT1})
- (\ref{GDNAlge1SPLT4}).

Next we imagine there is a sender who wishes to send an encrypted version of
these to a receiver.  The method of encryption is very simple.  The encrypted 
versions transmitted in the open have all their $\ell$-parameters and $\Hat 
\ell$-parameters set to zero.  From the examples we have worked out previously,
we know in some cases (with a relatively small amount of effort) the receiver can
set up calculations to reconstruct the encrypted matrices.  What the examples do
not show us is how general is this capability.  We assert understanding this problem
in its generality is equivalent to solving the adinkra network version of the auxiliary
field problem.  As cryptography is a very well developed topic, it may well be that
this alternate formulation of the problem can take advantage of some of this pre-existing
knowledge.

\section{Summary and Conclusion} 
$~~~$
The most important result of this work is the demonstration that given the information 
of an on-shell adinkra network it is possible by use of the Garden Algebra to {\em 
{derive}} a corresponding off-shell structure in which the on-shell one is embedded.  

The method we have introduced involves the introduction of a space of real parameters,
denoted by $\ell$'s, which are used to construct matrices that interpolate from a
description of an on-shell adinkra network to an off-shell one.  There may be an 
interesting mathematical question to pursue here.  If we think of the $\ell$'s as 
the coordinates of some space, then the solution to the ``Garden Algebra'' problem
for augmented on-shell L-matrices and R-matirces may be regarded as the search 
for the loci of points which simultaneously solve the conditions arising from the 
``Garden Algebra.''  This offers the possibility of attacking such problems from the 
point of view of real algebraic geometry.  Within the DFGHILM \cite{DFGHILMwbp} collaboration,
but in unpublished private discussions, it has long been recognized that for
some theories (with more than four colors), there exist the possibility that there
not only exist isolated points that satisfy the ``Garden Algebra'' conditions,
but entire surfaces.

Via adinkras and their adjacency matrices, the off-shell auxiliary field problem of 
supersymmetrical systems has been ``translated'' into more precise mathematical
questions.  The statement of these problems can be cast in the following form.
Begin with a set of $\cal N$ d${}_L$ $\times$ d${}_R$ set of L-matrices and
a set of $\cal N$ d${}_R$ $\times$ d${}_L$ set of R-matrices.  By the augmentation
process described in the last chapter, these can be enlarged to be 4$p$
$\times$ $4p$ matrices for some integer $p$.  Given an arbitrary set of the
initial d${}_L$ $\times$ d${}_R$ and d${}_R$ $\times$ d${}_L$ matrices,
is it possible to find augmentations that satisfy the conditions in  (\ref{GDNAlge1SPLT1})
and  (\ref{GDNAlge1SPLT2})?

We have two conjectures to make along these lines.

Conjecture \# 1

$~~~~~$ In the work of \cite{G-1}, the L-matrices and R-matrices of a
formulation of the \newline \indent 4D,
 $\cal N$ = 1 double tensor adinkra network were given and this system
does not  \newline \indent
possess an augmentation satisfying (\ref{GDNAlge1SPLT1}) and
(\ref{GDNAlge1SPLT2}) in an irreducible manner.

Conjecture \# 2

$~~~~~$ In the work of \cite{adnkN4SYM2}, the L-matrices and R-matrices of a
formulation of the \newline \indent 4D,
 $\cal N$ = 4 Maxwell adinkra network were given and this system
possesses an   \newline \indent augmentation satisfying
(\ref{GDNAlge1SPLT1})
and (\ref{GDNAlge1SPLT2})  in an irreducible manner.

With this work, we provided a proof of concept
that the (R), (AD), and (I) steps are all implementable in the context of
supersymmetrical field theories.  However, even if one is successful
in all of these, there remains a challenge that caution bids us to note.
The (O) operation denoting the dimensional enhancement of the adinkra
network world results to then be converted first back into 0-brane world
results and hence dimensionally enhanced back to a full Minkowskian
space construction is not yet guaranteed to us.  It could be that there 
exist some obstruction to carrying out this step even though of the
other steps of the RADIO proposal are successful.

Though we are mindful of this possibility, we are also optimistic as in
recent times, we have developed an understanding and powerful tools
(``Adinkra/Gamma Matrix Equations,'' ``Coxeter Group Orbit/Hodge Duality Relations,''
and ``Holoraumy'') \cite{permutadnk1,permutadnk2,adnkholor1,adnkholor2}
which strongly suggest the existence of invariants that can be used
to start from an adinkra network world description and recover a
corresponding 0-brane world description.  Once this is done, we believe the
step of dimensional enhancement or (O) ``oxidation'' should be straightforward.

 \begin{center}
\parbox{4in}{{\it ``An error does not become truth by reason of multiplied 
propagation, nor does truth become error because nobody sees it. Truth 
stands, even if there be no public support. It is self sustained.''}\,\,-\,\, M. K. Ghandi 
$~~~~~~~~~$}
 \parbox{4in}{
 $~~$}  
 \end{center}
 
  \noindent
{\bf Acknowledgements}\\[.1in] \indent
We would like to acknowledge Professors Kevin Iga, Tristan H\" ubsch, Kory
Stiffler, and Stefan Mendez-Diaz for conversations.  This work was partially supported 
by the National Science Foundation grant PHY-0354401 and in part by the University of Maryland Center for 
String \& Particle Theory.  Additional acknowledgment is given by M.\ Calkins, 
and D.\ E.\ A.\ Gates to the Center  for String and Particle Theory, as well as 
recognition for their participation in 2013 \& 2014 SSTPRS (Student Summer 
Theoretical Physics Research Session) programs.  The adinkras in this work
were drawn with the aid of  T.\ H\" ubsch.  Finally, SJG wishes to thank J.\ H.\ Schwarz
for the introduction to this problem and encouragement over decades.    

\

\end{document}